\documentclass[pdflatex,sn-mathphys-num]{sn-jnl}% Math and Physical Sciences Numbered Reference Style 
\usepackage{graphicx}%
\usepackage{multirow}%
\usepackage{amsmath,amssymb,amsfonts}%
\usepackage{amsthm}%
\usepackage{mathrsfs}%
\usepackage[title]{appendix}%
\usepackage{xcolor}%
\usepackage{textcomp}%
\usepackage{manyfoot}%
\usepackage{booktabs}%
\usepackage{algorithm}%
\usepackage{algorithmicx}%
\usepackage{algpseudocode}%
\usepackage{listings}%
\usepackage{xcolor}
\usepackage[d]{esvect}
\usepackage{siunitx}
\usepackage{physics}
\AtBeginDocument{\RenewCommandCopy\qty\SI}
\DeclareSIUnit\torr{torr}
\usepackage[normalem]{ulem}
\DeclareRobustCommand{\erase}{\bgroup\markoverwith{\textcolor{red}{\rule[.5ex]{2pt}{0.4pt}}}\ULon}

\usepackage{bm}

\theoremstyle{thmstyleone}%
\theoremstyle{thmstyletwo}%
\theoremstyle{thmstylethree}%

\begin{document} 

\title[\textcolor{blue}{Force-Based Reading and Writing of Individual Single-Atom Magnets}]{Force-Based Reading and Writing of Individual Single-Atom Magnets}


\author{\fnm{Yuuki} \sur{Adachi}}\email{y-adachi@g.ecc.u-tokyo.ac.jp}

\author{\fnm{Kazuki} \sur{Ueda}}\email{ueda-kazuki972@g.ecc.u-tokyo.ac.jp}

\author{\fnm{Yuuki} \sur{Yasui}}\email{yasui@k.u-tokyo.ac.jp}

\author*{\fnm{Yoshiaki} \sur{Sugimoto}}\email{ysugimoto@k.u-tokyo.ac.jp}

\affil{Department of Advanced Materials Science, The University of Tokyo, Kashiwa, Chiba 277-8561, Japan}

\abstract{The integration of single-atom bits enables the realization of the highest data-density memory. Reading and writing information to these bits through mechanical interactions opens the possibility of operating the magnetic devices with low heat generation and high density recording. To achieve this visionary goal, we demonstrate the use of magnetic exchange force microscopy to read and write the spin orientation of individual holmium adatoms on MgO thin films. The spin orientation of the holmium adatom is stabilized by the strong uniaxial anisotropy of the adsorption site and can be read out by measuring the exchange forces between the magnetic tip and the atom. The spin orientation can be written by approaching the tip closer to the holmium adatom. We explain this writing mechanism by the symmetry reduction of the adsorption site of the Ho adatom. These findings demonstrate the potential for information storage with minimal energy loss and pave the way for a new field of atomic-scale mechano-spintronics.}

\maketitle

\textbf{Main text:}
A hard disk drive, which is a representative nonvolatile memory device, is composed of two-state magnetic bits that store information \cite{moser2002magnetic,5389136}. Traditionally, reading information from these magnetic bits has been performed through magnetoresistive effects driven by electric currents, while writing information to these magnetic bits relies on electromagnetic induction generated by current-driven magnetic fields. However, the use of electric currents inevitably causes Joule heating \cite{makarov2012emerging}. In this study, we develop current-free methods for reading and writing information to the single-atom bits, thereby achieving low heat generation and high recording density. Specifically, we propose a method for reading and writing information to the single-atom bits using force.

\begin{figure*}
\centering
\includegraphics[width=1.0\linewidth] {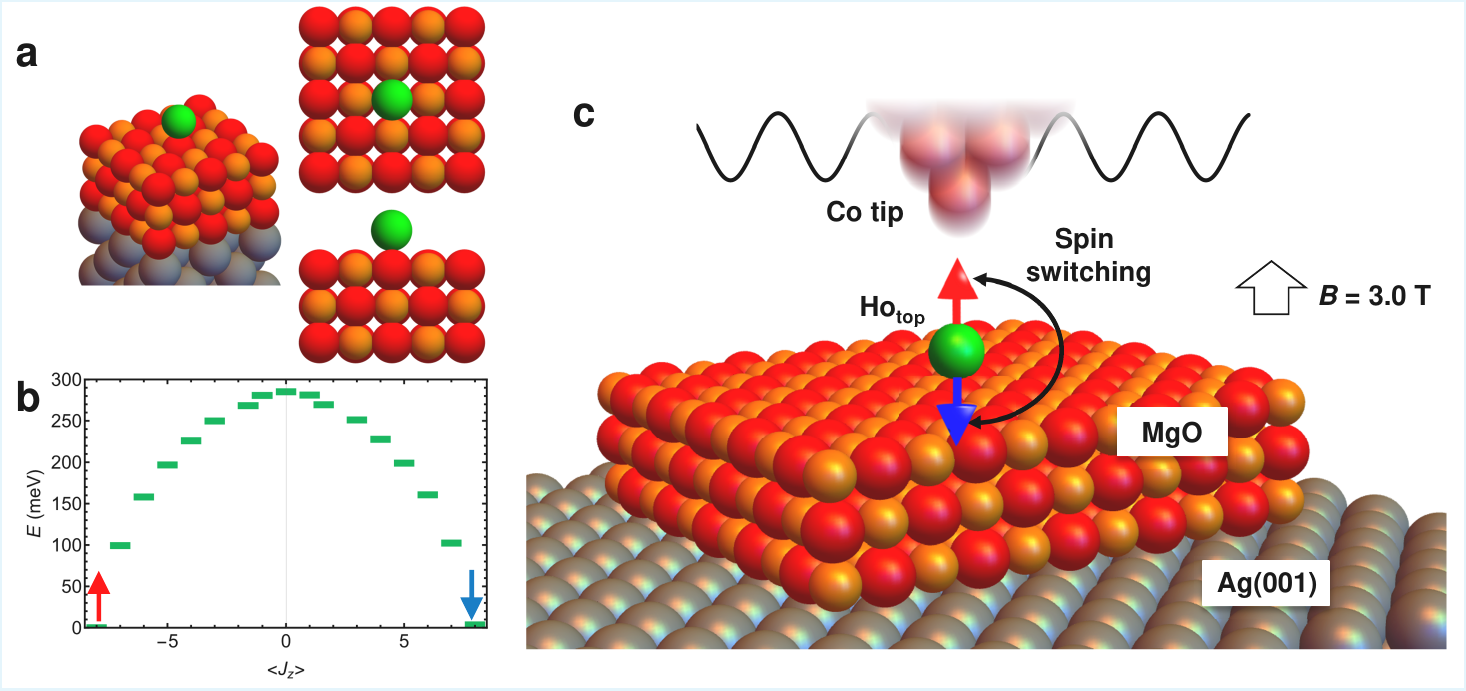}
\caption{\textbf{Energy diagram of holmium adatom on MgO and experimental set-up to read and write the magnetic states of Ho adatoms.}
(a) Three-dimensional views of the adsorption configuration of a Ho adatom at the top site in the high-symmetry $C_{4v}$ position on MgO/Ag(100), together with top and side views of the same configuration on MgO. Green ball: Ho atom, orange ball: Mg atom, red ball: O atom, gray ball: Ag atom. (b) Calculated eigenvalues of top-site Ho in high-symmetry $C_{4v}$ on MgO/Ag(100) at $B =  \SI{3.0}{\tesla}$. The large uniaxial crystal field, with only minor transverse components, suppresses efficient direct transitions between the ground and metastable states (Ho$\uparrow$ and Ho$\downarrow$). The red and blue arrows in (b) indicate the Ho$\uparrow$ and Ho$\downarrow$. (c) Schematic of the force-based reading and writing of single atom magnets. The Co tip mounted on a LER at a resonance frequency $f$$_{\rm{0}}$ of $\sim 1$ MHz with amplitude $A$ = \SI{65} pm.}\label{fig:Fig1}
\end{figure*}

Herein, we utilize holmium (Ho) adatoms on MgO as single-atom magnet, a system that is regarded as a benchmark for single-atom memory \cite{donati2021perspective,natterer2017reading,donati2016magnetic,sorokin2023impact,donati2020magnetic,natterer2018thermal,donati2020unconventional,thesisTobias} (see also Fig.S1 and Fig.S2 in the supporting information). 
 A Ho adatom adsorbed on the Oxygen top site (Ho$_{\rm{top}}$) has been experimentally characterized using X-ray absorption spectroscopy (XAS), X-ray magnetic circular dichroism (XMCD) and spin-polarized scanning tunneling microscopy (SP-STM) \cite{singha2021mapping,natterer2018thermal,donati2020unconventional}. As shown in Fig.~\ref{fig:Fig1}(a,b), Ho$_{\rm{top}}$ has a ligand field with $C_{4v}$ symmetry, which effectively suppresses direct transitions between the ground state and the metastable state (Ho↑ and Ho↓) due to the strong uniaxial anisotropy, and thus gives rise to a long-lived magnetic quantum state with two configurations, Ho↑ and Ho↓ (see also Section 3 in the Supporting Information) \cite{natterer2018thermal,thesisTobias,chen2023magnetic,natterer2017reading,forrester2019quantum}.
These properties make Ho$_{\rm{top}}$ adatoms promising candidates for the smallest stable magnetic bits. In previous SP-STM experiments, applying a bias voltage above $\sim$100 mV allowed current pulses to induce switching between Ho↑ and Ho↓ by overcoming the energy barrier ~\cite{natterer2017reading,natterer2018thermal}. Compared to Ho$_{\rm{top}}$, Ho$_{\rm{bridge}}$ is located in a crystal field with $C_{2v}$ symmetry, which represents a lower-symmetry environment (see Fig. S3). As depicted in Fig. S3b, this reduced symmetry leads to strongly mixed quantum states even under an applied magnetic field of 3.0 T, thereby shortening the magnetic lifetime \cite{thesisTobias}.

 To read and write the spin orientation of a Ho adatom adsorbed on an MgO surface, we employed magnetic exchange force microscopy (MExFM) using a length extension resonator (LER) operated in the frequency modulation mode \cite{kaiser2007magnetic,schmidt2011Quantitative,hauptmann2020quantifying,hauptmann2017sensing,pielmeier2013spin,adachi2025probing}. 
 Fig.~\ref{fig:Fig1}c illustrates the experimental setup. A tungsten tip functionalized with cobalt (Co) atoms at its apex was mounted on a LER, oscillated at its resonance frequency. It was then used to probe a Ho adatom under near-zero bias voltage at \SI{4.5}{\kelvin} under an external magnetic field of \SI{3.0}{\tesla} (see Fig. S4 and Fig. S5 in the Supporting Information for the preparation of the Co tip). The magnetization of the Co tip aligns with an external magnetic field in an upward direction due to the superparamagnetic nature of the Co cluster \cite{natterer2018thermal}.

\begin{figure*}
\vspace{-25mm}
\centering
\includegraphics[width=0.6 \linewidth] {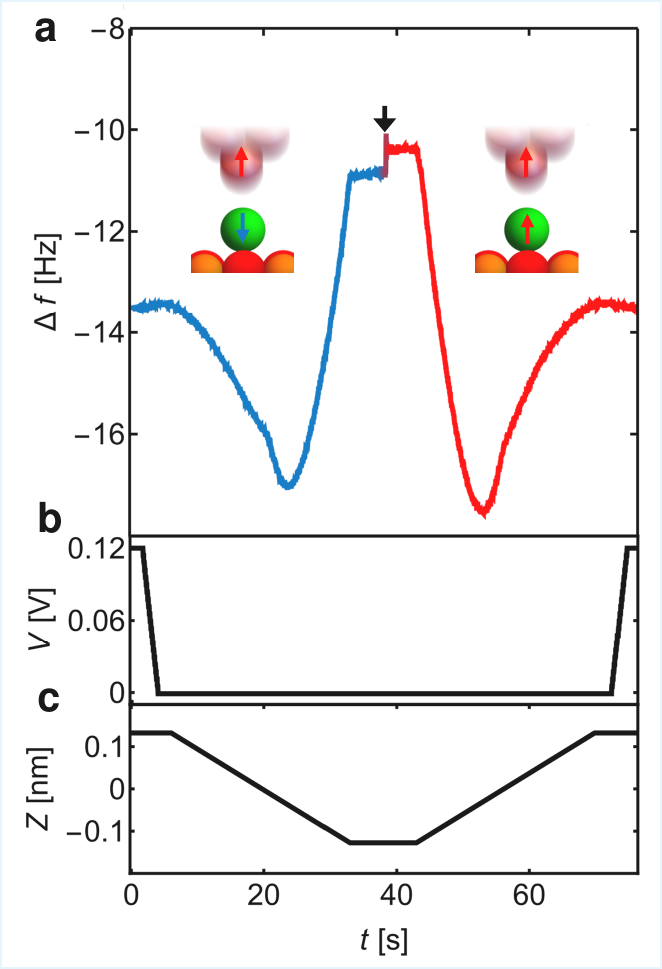}
\caption{\textbf{Reading and writing of Ho spin on MgO using MExFM.}
(a--c) $\Delta f(t)$ spectra measured on top of Ho$_{\rm{top}}$ while varying the bias voltage and the tip height. (a) Time evolution of $\Delta f$, (b) applied bias voltage, and (c) tip–sample distance. The blue and red in (a) indicate the Ho$\downarrow$ and Ho$\uparrow$. At \SI{33}{\second} $\leq t \leq$ \SI{43}{\second}, the transition from the Ho$\downarrow$ to Ho$\uparrow$ state can be detected by a sudden jump in $\Delta f(t)$, marked by the black arrow in (a).%MgO = 2 ML
}\label{fig:Fig2}
\end{figure*}

Here, we present the experimental results of force-based reading and writing Ho↓ and Ho↑. Fig.~\ref{fig:Fig2}a shows a typical frequency shift ($\Delta$\textit{$f$}) as a function of time, measured on top of Ho$_{\rm{top}}$ while varying the bias voltage (Fig.~\ref{fig:Fig2}b) and the tip--sample distances (Fig.~\ref{fig:Fig2}c).  First, the lateral position of the tip was fixed above the center of Ho$_{\rm{top}}$, and its spin state was stabilized in the desired configuration (in this case, Ho↓) by lowering the bias voltage from \textit{V} = \SI{+120}{mV} to \textit{V} = \SI{+200}{\micro\volt}, well below the threshold for current-induced spin switching (\SI{0}{\second} $\leq t \leq$ \SI{4}{\second}). Once set, the Ho↓ state was measured via $\Delta f$ during the tip approach (\SI{6}{\second} $ \leq t \leq $ \SI{33}{\second}). To switch the spin state from Ho↓ to Ho↑, the tip was brought to a specific distance (\textit{z} = $-0.13$ nm), exceeding the threshold required to induce spin switching (\textit{z} = $0.00$ nm is the point-contact distances, see also Supporting information section 1). The spin state of the Ho$_{\rm{top}}$ was then probed at this distance with a fixed probe time of 10 s (\SI{33}{\second} $ \leq t \leq $ \SI{43}{\second}), and the transition from Ho↓ to Ho↑ was detected as a sudden jump in $\Delta f$, indicated by the black arrow in Fig.~\ref{fig:Fig2}a. Afterward, the Ho↑ was measured from $\Delta f$ by retracting the tip to its original tip--sample distances (\SI{43}{\second} $ \leq t \leq $ \SI{70}{\second}). Finally, the bias voltage was restored from \textit{V} = \SI{+200}{\micro\volt} to its original value of \textit{V} = \SI{+120}{mV} (\SI{72}{\second} $\leq t \leq$ \SI{76}{\second}; see also Fig. S6 for the tunneling current simultaneously recorded with $\Delta f$ in Fig.~\ref{fig:Fig2}a). In Fig.~\ref{fig:Fig2}a, the minimum of $\Delta f$ obtained on top of Ho↑ is smaller than that for Ho↓, demonstrating that the two spin states can be successfully read out by MExFM, and that the Ho spin can be written from Ho↓ to Ho↑ by approaching the tip.

\begin{figure*}
\centering
\includegraphics[width=0.6\linewidth] {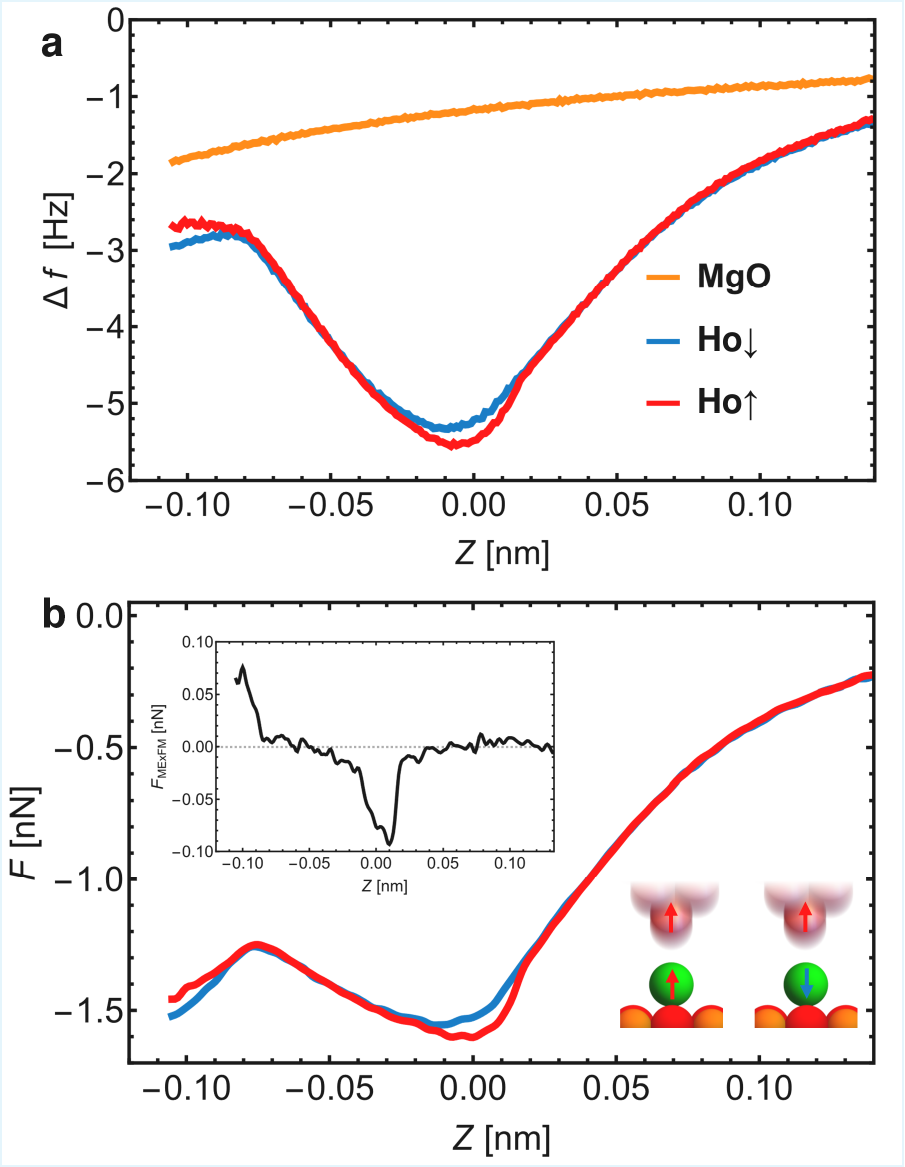}
\caption{\textbf{Probing Ho$\uparrow$ and Ho$\downarrow$ using MExFM.}
(a) Frequency shift obtained on top of the Ho$\uparrow$ ($\Delta$\textit{$f$}$_{\rm{Ho\uparrow}}$(\textit{$z$}), red solid curve),  Ho$\downarrow$ ($\Delta$\textit{$f$}$_{\rm{Ho\downarrow}}$(\textit{$z$}), blue solid curve) and MgO ($\Delta f_{\mathrm{MgO}}(z)$, orange solid curve). Measurement conditions: $V = \SI{+ 200}{\micro\volt}$. 
(b) Short-range forces obtained on top of Ho$\uparrow$ (\textit{$F$}$_{\rm{Ho\uparrow}}$(\textit{$z$}), red solid curve) and Ho$\downarrow$ (\textit{$F$}$_{\rm{Ho\downarrow}}$(\textit{$z$}), blue solid curve).  
Inset in (b) shows magnetic exchange force \textit{$F$}$_{\rm{MExFM}}$(\textit{$z$}) obtained on top of the Ho adatom. The gray dotted line is a guide for the eye, indicating \textit{$F$}$_{\rm{MExFM}}$(\textit{$z$}) = 0.
}\label{fig:Fig3}
\end{figure*}

  To discuss the reading mechanism of Ho↑ and Ho↓, here we show the short-range force and magnetic exchange force recorded on top of Ho$_{\rm{top}}$ (see Fig. S7 and Fig. S8 for full data sets and dissipation). Fig.~\ref{fig:Fig3}a shows the $\Delta$\textit{$f$} as a function of tip--sample distance ($\Delta$\textit{$f$}(\textit{$z$})), recorded on top of Ho↑, Ho↓ and MgO ($\Delta$\textit{$f$}$_{\rm{Ho\uparrow}}$($z$),  $\Delta$\textit{$f$}$_{\rm{Ho\downarrow}}$($z$) and $\Delta$\textit{$f$}$_{\rm{MgO}}$(\textit{$z$})). Both $\Delta$\textit{$f$}$_{\rm{Ho\uparrow}}$($z$) and $\Delta$\textit{$f$}$_{\rm{Ho\downarrow}}$($z$) include a long-range component from the MgO substrate. Therefore, $\Delta$\textit{$f$}$_{\rm{MgO}}$(\textit{$z$}) was subtracted from $\Delta$\textit{$f$}$_{\rm{Ho\uparrow}}$(\textit{$z$}) and $\Delta$\textit{$f$}$_{\rm{Ho\downarrow}}$(\textit{$z$}) to eliminate the background component. As shown in Fig.~\ref{fig:Fig3}b, the short-range force on Ho↑ and Ho↓ (\textit{$F$}$_{\rm{Ho\uparrow}}$(\textit{$z$}) and \textit{$F$}$_{\rm{Ho\downarrow}}$(\textit{$z$})) were calculated from the background subtracted $\Delta$\textit{$f$}(\textit{$z$}) \cite{sader2004accurate}. In Fig.~\ref{fig:Fig3}b, as the tip approaches the Ho↑ (Ho↓) adatom, \textit{$F$}$_{\rm{Ho\uparrow}}$($z$) (\textit{$F$}$_{\rm{Ho\downarrow}}$($z$)) exhibits \textit{$F$}$_{\rm{Ho\uparrow}}$($z = 0.00$ nm) $= -1.60$ nN (\textit{$F$}$_{\rm{Ho\downarrow}}$($z = 0.00$ nm) $= -1.55$ nN), indicating ferromagnetic coupling between the Co tip and the Ho adatom at this distance. Reducing $z$ further decreases the attractive force to \textit{$F$}$_{\rm{Ho\uparrow}}$($z = -0.08$ nm) $= -1.25$ nN (\textit{$F$}$_{\rm{Ho\downarrow}}$($z = -0.08$ nm) $= -1.25$ nN). As $z$ is reduced even more, the attraction increases, reaching \textit{$F$}$_{\rm{Ho\uparrow}}$($z = -0.10$ nm) $= -1.40$ nN (\textit{$F$}$_{\rm{Ho\downarrow}}$($z = -0.10$ nm) $= -1.50$ nN), indicating antiferromagnetic coupling at this distance. The inset in Fig.~\ref{fig:Fig3}b shows the magnetic exchange force, $F_{\rm MExFM}(z)$, derived by subtracting $F_{\rm Ho\downarrow}(z)$ from $F_{\rm Ho\uparrow}(z)$. As the tip approaches, a transition from ferromagnetic to antiferromagnetic coupling can be observed (ferromagnetic: $F_{\rm MExFM}(z) < 0$ and antiferromagnetic: $F_{\rm MExFM}(z) > 0$).

% Fig.3c further reveals deviations from a strictly exponential exchange force evolution, which are present in the range before a point contact is formed. This deviation, estimated to be about 10~pm, is attributed to atomic relaxations in the junction, a phenomenon often observed in STM experiments on both nonmagnetic and magnetic surfaces as well as on adatoms.

 The ferromagnetic coupling between the highly localized 4$f$ electrons in the Ho adatom and the 3$d$ electrons in the Co tip can be explained by two contributions: first, an intra-atomic ferromagnetic coupling between the 4$f$ and 5$d$ (or 6$s$) spins within the Ho adatom, and second, an inter-atomic ferromagnetic coupling between the 5$d$ (or 6$s$) electrons of Ho and the 3$d$ electrons of the Co atom \cite{pivetta2020measuring,natterer2018thermal,singha2018spin}. The transition from ferromagnetic to antiferromagnetic coupling is
 reported for the interaction between the 5$d$ electrons of Ta and the 3$d$ electrons of Fe \cite{wieser2013theoretical}. 
 
\begin{figure*}
\centering
\includegraphics[width=0.8 \linewidth] {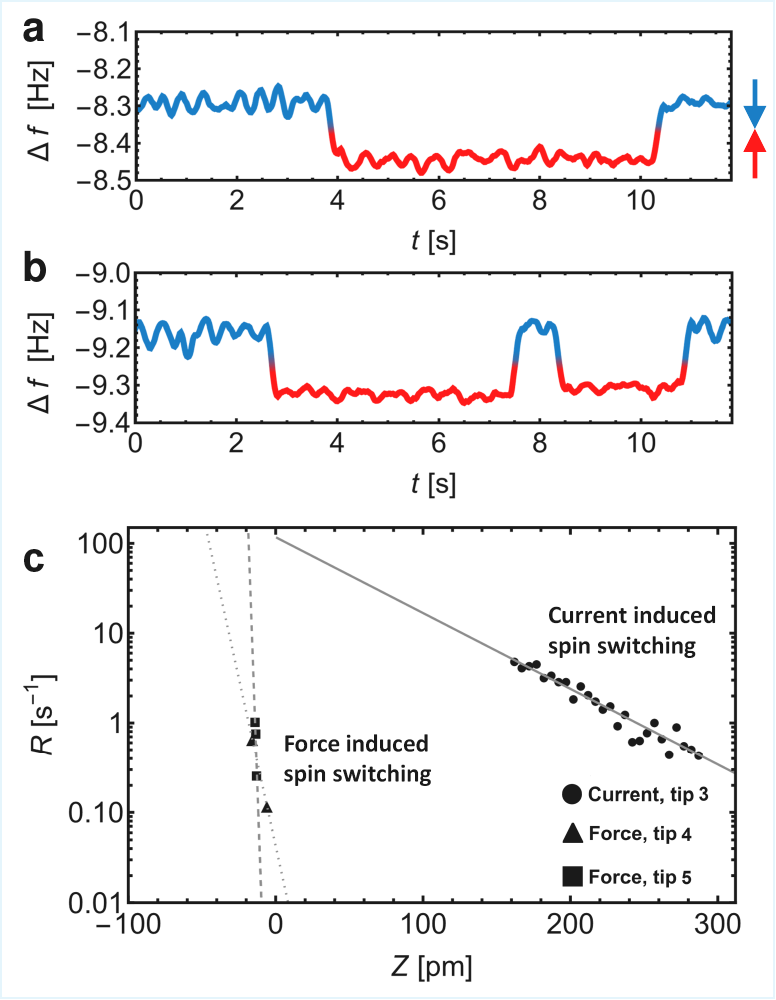}
\caption{\textbf{Switching between Ho$\uparrow$ and Ho$\downarrow$.}
(a,b) Telegraph signal due to the force-induced magnetic switching between Ho$\uparrow$ and Ho$\downarrow$ by changing the tip height. The blue and red indicate the Ho$\downarrow$ and Ho$\uparrow$.
Measurement conditions: constant-height mode, \textit{$V$} $=$ \SI{+ 1.0}{\milli\volt}, (a) \textit{$z$} = \SI{-6.0}{\pm}; (b) \textit{$z$} = \SI{-16.0}{\pm}.  (c) Spin switching rate as a function of tip--sample distances. The exponential fits are represented by the solid, dashed and dotted lines. Measurement conditions: constant height mode, \textit{$V$} = \SI{1.0}{\mV} for force induced spin switching and \textit{$V$} = \SI{150}{\mV} for current induced spin switching.}\label{fig:Fig4}
\end{figure*}

As we discussed in Figs.~\ref{fig:Fig2}(a--c), the Ho spin can be switched from Ho$\downarrow$ to Ho$\uparrow$ by approaching the tip. In Figs.~\ref{fig:Fig4}(a,b), we further demonstrate bidirectional switching between Ho$\downarrow$ and Ho$\uparrow$ induced by the tip approach. Firstly, the tip was brought to the center of Ho$_{\rm{top}}$, and the bias voltage was set to \textit{V} = \SI{+200}{\micro\volt} to avoid spin switching induced by the tunneling current. Then, the tip–sample distance was adjusted to values exceeding the threshold required to induce spin switching via tip approach. As shown in Figs.~\ref{fig:Fig4}(a,b), the spin switching was monitored in real time by recording the $\Delta f$ while keeping the tip height constant. Telegraph noise between the two states was observed, indicating bidirectional spin switching between Ho↑ and Ho↓. Therefore, due to the bidirectional spin switching, we can control the spin not only from Ho↓ to Ho↑ (as demonstrated in Figs.~\ref{fig:Fig2}(a--c)) but also from Ho↑ to Ho↓, as shown in Fig.~S9. Moreover, the observation of the bidirectional spin switching rules out exchange forces as the driving mechanism for spin switching \cite{schmidt2012magnetization}. Fig.~\ref{fig:Fig4}c summarizes the switching rates between Ho↑ and Ho↓ with the results of spin switching induced by the tunneling current (see also Fig. S10).  In Fig.~\ref{fig:Fig4}c, the spin switching induced by tip approach decays more rapidly along with distance than that induced by the tunneling current. These results suggest that force-based spin switching enables a more spatially localized control of the spin orientation in the Ho adatom.

To discuss the writing mechanism of Ho↑ and Ho↓, the tip was positioned at the center of Ho$_{\rm top}$ and approached closer than the spin-switching distance. As shown in Fig. S11, this approach induced a lateral displacement of Ho$_{\rm top}$ to the bridge site, resulting in the formation of Ho$_{\rm bridge}$. These results demonstrate that lateral displacement from Ho$_{\rm top}$ to Ho$_{\rm bridge}$ can be induced even when the tip approaches the center of Ho$_{\rm top}$ at a specific tip–sample distance. Because the Co tip has an asymmetric shape (see Fig. S12(a--d)), multiple Co atoms are expected to come into contact with the Ho adatom during relaxation, thereby inducing lateral displacement. Notably, once the Ho adatom relocates to the Ho$_{\rm bridge}$, it rarely returns to Ho$_{\rm top}$.

The lateral displacement from Ho$_{\rm top}$ to Ho$_{\rm bridge}$, in turn, strongly influences the spin state of the Ho adatom. Specifically, as shown in Fig.~\ref{fig:Fig1}b and Fig. S3b, transitions between Ho$\uparrow$ and Ho$\downarrow$ are suppressed for Ho$_{\rm top}$, whereas the reduced-symmetry Ho$_{\rm bridge}$ exhibits strongly mixed quantum states, enabling direct transitions between Ho$\uparrow$ and Ho$\downarrow$ even under an applied magnetic field of 3.0 T (see also Section 3 in the Supporting Information)\cite{thesisTobias}. The importance of crystal-field symmetry for magnetic stability has been widely reported in other systems \cite{hubner2014symmetry,miyamachi2013stabilizing,sorokin2023impact}. We therefore propose that spin switching between Ho↑ and Ho↓ occurs when Ho$_{\rm top}$ moves toward Ho$_{\rm bridge}$ but does not fully reach it, due to the force exerted by the Co tip. This is further confirmed as in Fig. S13, which shows that the spin switching distance varies depending on the tip shape, but spin switching always occurs at distances shorter than the point-contact distance. Although the spin switching distance depends on the tip shape, once an appropriate tip is prepared, the spin orientation and switching rate can be controlled by adjusting the tip–sample distance, as demonstrated in Fig.~\ref{fig:Fig2}, Fig. S9, and Figs.~\ref{fig:Fig4}(a--c), thereby enabling controlled writing of a single-atom magnet using force.

\begin{figure*}
\vspace{-25mm}
\centering
\includegraphics[width=0.8 \linewidth] {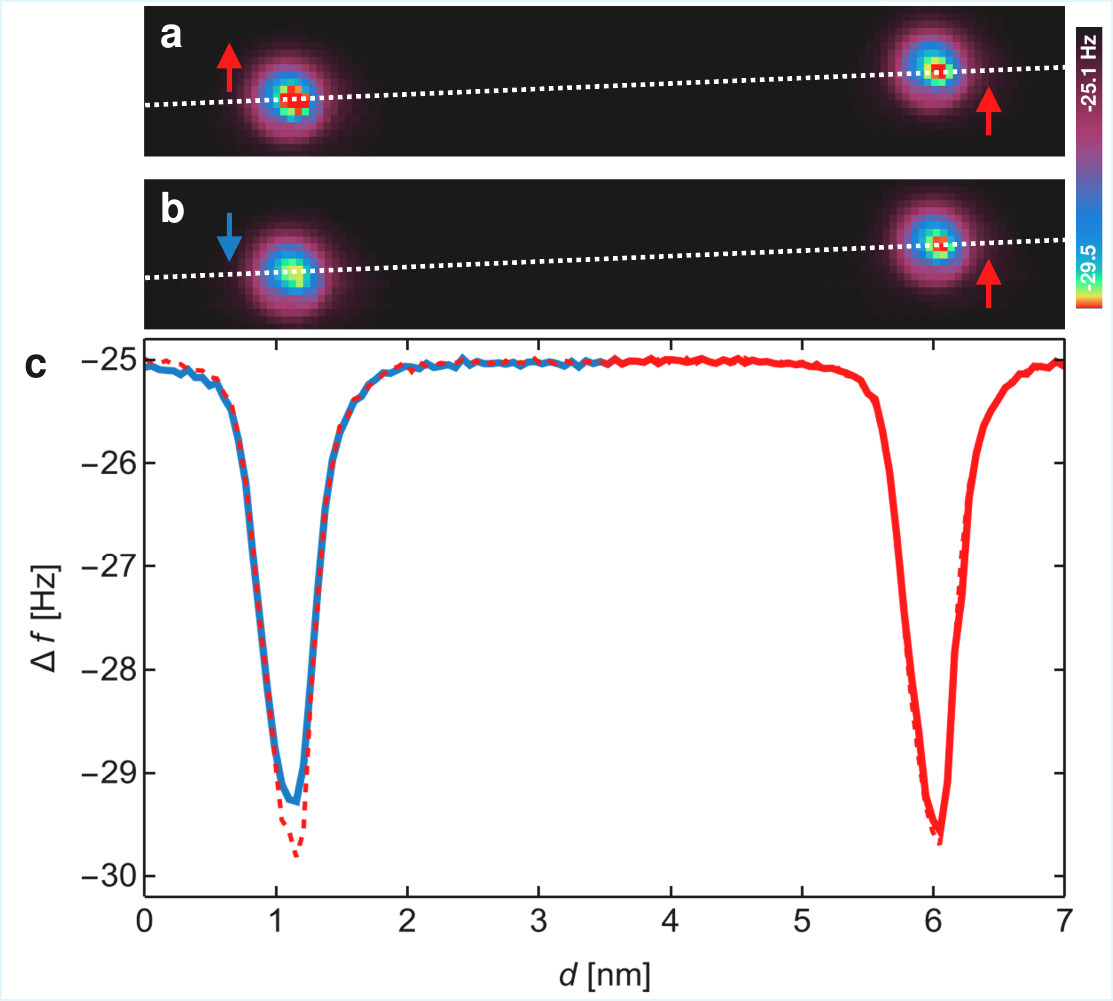}
\caption{\textbf{Imaging Ho$\uparrow$ and Ho$\downarrow$ using MExFM.}
 (a) $\Delta f$ image of two Ho adatoms, both Ho adatom in the Ho$\uparrow$. Imaging parameters: constant height mode, \textit{$V$} $=$ \SI{+ 200}{\micro\volt}. (b) $\Delta f$ image of the same area in (a), after the left Ho adatom was manipulated from Ho$\uparrow$ to Ho$\downarrow$. Imaging parameters: constant height mode, \textit{$V$} $=$ \SI{+ 200}{\micro\volt}. 
(a) and (b) were obtained at the same tip height. (c) Line profiles obtained above the Ho adatoms by the dotted curve for (a) and the solid curve for (b). The blue and red indicate the Ho$\downarrow$ and Ho$\uparrow$. The positions of the line profiles are indicated by the dotted lines in (a) and (b). %MgO = 2 ML
}\label{fig:Fig5}
\end{figure*}

Beyond merely reading Ho↓ and Ho↑ through its spectroscopy capabilities, MExFM enables imaging of Ho↓ and Ho↑. As shown in Figs.~\ref{fig:Fig5}(a,b), this is achieved by scanning the tip horizontally at a constant height while recording $\Delta f$. The tip–sample distance is set to be approximately 20 pm larger than the point-contact distance to avoid spin switching. In Fig.~\ref{fig:Fig5}a, both Ho adatoms appear in the Ho↑ state. To demonstrate the spin-readout capability, we switched the spin state of the left Ho adatom from Ho↑ to Ho↓ and imaged the same area again. In Fig.~\ref{fig:Fig5}b, the left Ho adatom appears in the Ho↓ state, whereas the right one remains in the Ho↑ state (see also Fig.~\ref{fig:Fig5}c), confirming the successful readout of the spin orientation of Ho adatoms in both spin configurations using MExFM.

 The method of controlling the spin by manipulating the adsorption-site symmetry, as proposed in this study, is not specific to our system. This opens new possibilities for manipulating spin states via the surrounding atomic environment. Spin detection and manipulation of 4$f$-electron systems by a nondissipative force, unlike electric currents, is expected to lead to the realization of long spin coherence times, which are critical for quantum
information processing \cite{sellies2023single,czap2025direct,reale2024electrically}.



\clearpage




\clearpage

\clearpage

\clearpage

\clearpage

\clearpage
\section*{Author contributions}
Conceptualization: Y.S., Y.Y., Y.A. AFM and STM observation: Y.A., K.U. Writing original draft: Y.A. Writing review and editing: Y.A., Y.Y., Y.S.

\section*{Conflicts of interest}
There are no conflicts to declare.

\section*{Data availability}
The data supporting this article have been included as part of the ESI.

\section*{Acknowledgements}
We thank T. Ozaki, M. Fukuda for theoretical discussions, and E. Kazuma and M. Lee for discussions on sample preparation of the MgO thin film. This work was supported by the JST FOREST Program (Grant No. JPMJFR203J) and JSPS KAKENHI (Grant Nos. 25K22215, 25K17946, 25K17933, JP24H01175 and JP23K13656). Y.S. acknowledges the support of the Asahi Glass Foundation and the Murata Science Foundation. Y.A. acknowledges the support of the Technology Foundation.

\clearpage
% \end{thebibliography}
\bibliography{Referencemain}

%%
%% End of file `elsarticle-template-num.tex'.

% Produces the bibliography via BibTeX.
\end{document}

\title[Supporting information: 
Force-Based Reading and Writing of Individual Single-Atom Magnets]{Supporting information: 
Force-Based Reading and Writing of Individual Single-Atom Magnets}

\author{\fnm{Yuuki} \sur{Adachi}}\email{y-adachi@g.ecc.u-tokyo.ac.jp}

\author{\fnm{Kazuki} \sur{Ueda}}\email{ueda-kazuki972@g.ecc.u-tokyo.ac.jp}

\author{\fnm{Yuuki} \sur{Yasui}}\email{yasui@k.u-tokyo.ac.jp}

\author*{\fnm{Yoshiaki} \sur{Sugimoto}}\email{ysugimoto@k.u-tokyo.ac.jp}

\affil{Department of Advanced Materials Science, The University of Tokyo, Kashiwa, Chiba 277-8561, Japan}

\maketitle

\section{Materials and method}
The experiments were conducted using a custom-built scanning probe microscope operating under ultra-high vacuum conditions at \SI{4.5}{\kelvin} \cite{sugimoto2019force}. A commercial quartz length-extension resonator (LER), known as a Kolibri sensor (SPECS) and equipped with a tungsten (W) tip, was utilized. The microscope is a combined AFM/STM system capable of applying magnetic fields of up to \SI{11}{\tesla} perpendicular to the sample surface. The experiments were carried out with magnetic fields at \SI{3.0}{\tesla}. For differential conductance (d\textit{$I$}/d\textit{$V$}) spectroscopy, we employed a lock-in technique with a modulated sample bias voltage of \SI{2.0}{\mV} at \SI{617}{\Hz}. $I_{\rm average}$ indicates the time-averaged tunneling current measured by oscillating the tip at an amplitude of 65~pm. $I$ indicates the static tunneling current measured without oscillation. We have used several different Co tips for spin switching. To make it easier to distinguish them, we assigned the same numerical label to the same tip, as shown in Table 1.

\begin{figure*}
\centering
\includegraphics[width=1.0 \linewidth] {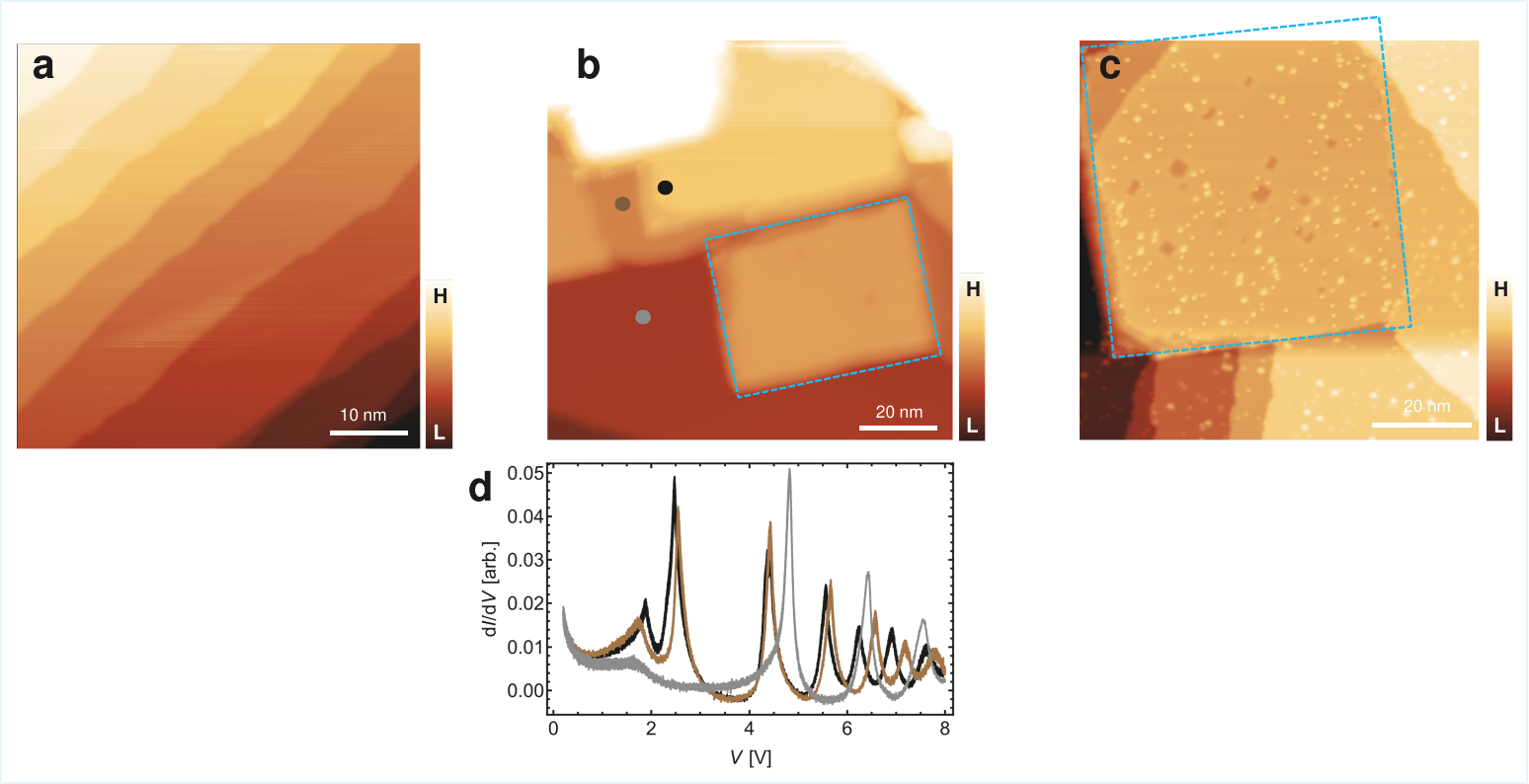}
\caption{\textbf{STM images of Ag(001) and MgO/Ag(001).}
(a--c) STM topography images of Ag(001), MgO/Ag(001), and coadsorbed Ho and Co on MgO/Ag(001). The blue dotted rectangles indicate the MgO thin film. Imaging conditions: constant-current mode. (a) \textit{$I$} = \SI{1.0}{\nA}, \textit{$V$} = \SI{200}{\mV}; (b) \textit{$I$} = \SI{1.0}{\nA}, \textit{$V$} = \SI{200}{\mV}; (c) \textit{$I$} = \SI{200}{\pA}, \textit{$V$} = \SI{400}{\mV}. (d) Field emission spectra measured on Ag(001) (gray) and MgO/Ag(001) (brown and black) in constant-current mode (\textit{$I$} = \SI{1}{\nA}). The spectroscopic locations are indicated by different colors in (b).}\label{fig:FigS1}
\end{figure*}

The Ag(001) surface was cleaned by repeated cycles of Ar$^{+}$ sputtering followed by annealing at \SI{600}{\degreeCelsius}. The quality of Ag(001) was confirmed by STM imaging (Fig.~\ref{fig:FigS1}a). After confirming the clean Ag(001), Mg was deposited in an oxygen background pressure of \SI{2.0E-7}{\torr} at a rate of approximately 0.2 ML/min while the Ag(001) crystal was maintained at \SI{600}{\degreeCelsius}. The temperature was then slowly decreased to room temperature over a period of approximately 45 minutes \cite{pal2014morphology,pal2014growing,schintke2001insulator,schintke2004insulators,baumann2014measuring}. The quality of the MgO thin film (\SI{}{\geq} 2 ML) was confirmed by STM imaging and field-emission measurements (Fig.~\ref{fig:FigS1}b and Fig.~\ref{fig:FigS1}d) \cite{baumann2014measuring}. All the data are obtained on 3 ML films. A high-purity (\SI{99.9}{\percent}) Ho rod was cleaned by filing the oxidized surface layer until a shiny metallic surface was exposed. To minimize exposure to ambient conditions, the Ho rod was immediately transferred into the ultra-high vacuum chamber. Both holmium and cobalt atoms were deposited onto the MgO/Ag(001) at \SI{4.5}{\kelvin} (Fig.~\ref{fig:FigS1}c). 

 $z = 0$ is defined as the tip--atom distance, measured from the point conductance of Ho adatom \cite{paul2017control,ternes2011interplay}. When a Co tip was used, the point of conductance of Ho↓ was taken as the reference. We used point conductance \textit{$G$}$_{\rm{0}}$ = \SI{7.748E-5}{\siemens}.
 
\begin{table}[htbp]
\centering
\caption{Summary of the tips used for each figure in the main text and supplementary information. The same numerical label indicates the same tip.}
\begin{tabular}{lc}
\hline
Figure & Tip Used \\
\hline
Fig.~2   & Tip 1 \\
Fig.~3   & Tip 2 \\
Fig.~4   & Tips (3--5) \\
Fig.~5   & Tip 6 \\
Fig.~S1  & Tip 7 \\
Fig.~S2  & Tip 8 \\
Fig.~S4  & Tip 9 \\
Fig.~S5  & Tip 10 \\
Fig.~S6  & Tip 1 \\
Fig.~S7  & Tip 2 \\
Fig.~S8  & Tip 11 \\
Fig.~S9  & Tip 12 \\
Fig.~S10 & Tip 3 \\
Fig.~S11 & Tip 13 \\
Fig.~S12 & Tip 14 \\
Fig.~S13 & Tip 1, Tip 4, Tip 12 and Tips (15--21)\\
\hline
\end{tabular}
\end{table}

\section{Observation of coadsorbed Ho and Co atoms on MgO/Ag(001)}
\begin{figure*}
\centering
\includegraphics[width=1.0 \linewidth] {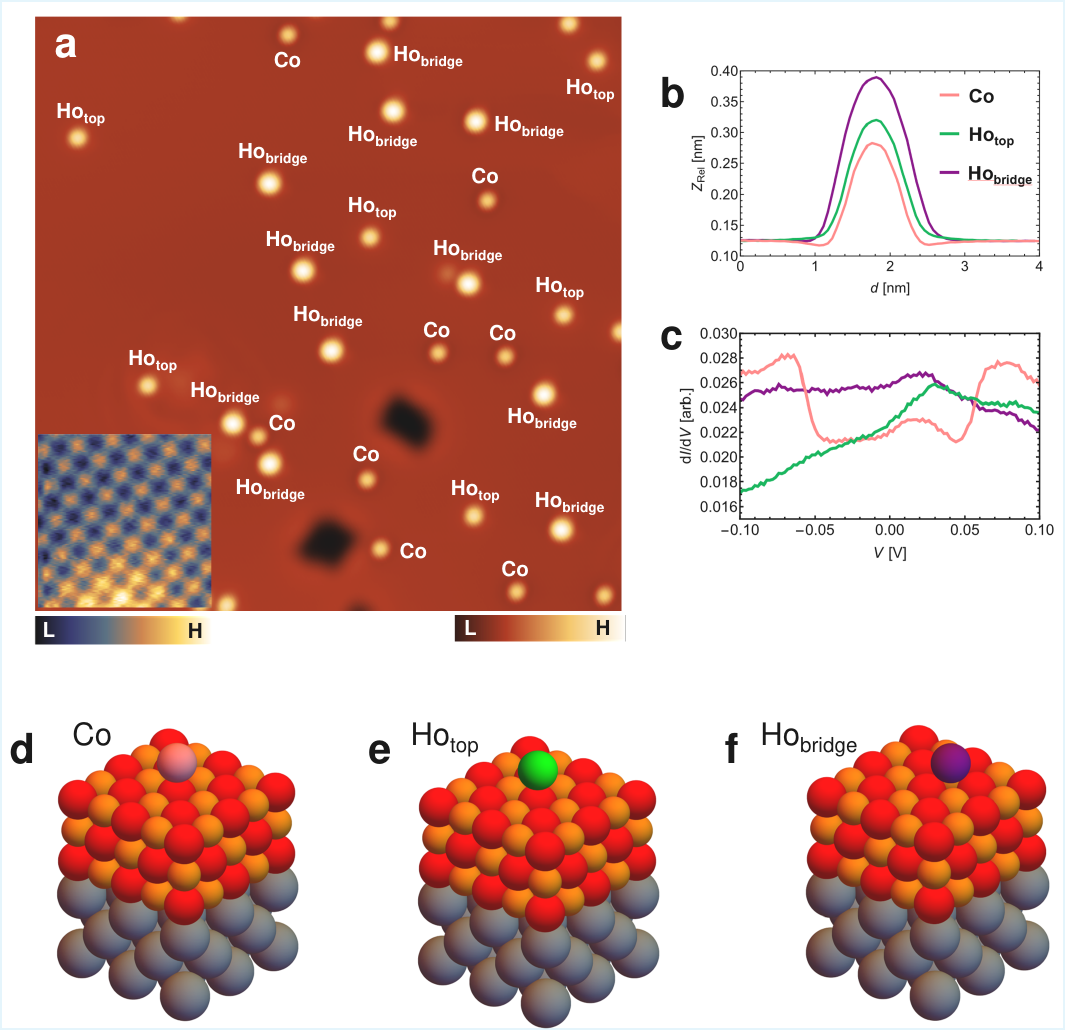}
\caption{\textbf{Identification of Co, Ho$_{\rm{top}}$ and Ho$_{\rm{bridge}}$}
 (a) STM topography images of coadsorbed Ho and Co on MgO/Ag(001). The inset shows the atomically resolved $\Delta$\textit{$f$} image of MgO thin film. Imaging condition for the STM topography image: constant current mode. \textit{$I$} = \SI{+ 1.0}{\nA}, \textit{$V$} = \SI{200}{\mV},  \textit{$B$} = \SI{+ 3.0}{\tesla}. Imaging condition for the inset image: constant height mode. \textit{$V$} = \SI{6.2}{\mV},  \textit{$B$} = \SI{+ 3.0}{\tesla}. (b) Line profile measured above Co, Ho$_{\rm{top}}$ and Ho$_{\rm{bridge}}$. (c) d\textit{$I$}/d\textit{$V$} spectrum measured on top of Co, Ho$_{\rm{top}}$ and Ho$_{\rm{bridge}}$. Same color code as (b). (d) Schematic of Co adatom located on MgO/Ag(001). (e,f) Schematic of Ho adatom located on top and bridge site. Pink ball: Co atom, green ball: Ho atom, purple ball: Ho atom, orange ball: Mg atom, red ball: O atom, gray ball: Ag atom in (d--f).}\label{fig:FigS2}
\end{figure*}

Fig.~\ref{fig:FigS2}a shows an STM image of coadsorbed Ho and Co atoms on MgO/Ag(001). Different types of stable bright features are observed on top of the MgO surface. Figs.~\ref{fig:FigS2}b and ~\ref{fig:FigS2}c show line profiles and d\textit{$I$}/d\textit{$V$} spectra measured on top of them. Ho$_{\rm top}$ and Ho$_{\rm bridge}$ are distinguished by their different apparent heights (Fig.~\ref{fig:FigS2}b). Moreover, in both cases, no inelastic conductance steps were observed  (Fig.~\ref{fig:FigS2}c) \cite{singha2018spin}. Co adatoms are identified by their d\textit{$I$}/d\textit{$V$} steps at 58 meV, reflecting their high magnetic anisotropy (Fig.~\ref{fig:FigS2}c) \cite{rau2014reaching, singha2018spin}. Schematics of these adatoms are shown in Figs.~\ref{fig:FigS2}(d–-f).

\section{The effective spin Hamiltonian analysis}
 We model the energy diagram of Ho adatom using an effective spin Hamiltonian of the form:

\begin{equation}
H = H_{\rm cf} + H_{\rm Z}
   \tag{S1}
\end{equation}
with $H_{\rm cf}$ describing the effects of the crystal field, and $H_{\rm Z}$ describing the effects of the external magnetic field. We use this effective spin Hamiltonian to describe the splitting of the lowest multiplet of magnetic states. For the Ho adatom in the $4f^{10}$ configuration, the lowest multiplet consists of states with a total magnetic moment $J = 8$ \cite{natterer2018thermal,forrester2019quantum}. This quantum number determines the corresponding multiplicity of the states, $2J+1 = 17$. 

For the Ho atoms adsorbed on the top site (Ho$_{\rm top}$), we include the Stevens operators permitted by the four-fold symmetry: 
\begin{equation}
H_{\mathrm{cf}\, C_{4v}} =
B_2^0 O_2^0 + 
B_4^0 O_4^0 + 
B_4^4 O_4^4 + 
B_6^0 O_6^0 + 
B_6^4 O_6^4
   \tag{S2}
\end{equation}

 On the other hand, for the Ho atoms adsorbed on the bridge site (Ho$_{\rm bridge}$), we include the Stevens operators permitted by the two-fold symmetry:

\begin{equation}
H_{\mathrm{cf}\, C_{2v}}=
B_2^0 O_2^0 + 
B_4^0 O_4^0 + 
B_6^0 O_6^0 + 
B_2^2 O_2^2 +
B_4^2 O_4^2 + 
B_4^4 O_4^4 + 
B_6^2 O_6^2 + 
B_6^4 O_6^4 +
B_6^6 O_6^6  
   \tag{S3}
\end{equation}

The values of the $B_n^k$ coefficients determine the zero-field splitting of the magnetic states. $O_n^k$ is the Stevens operators. To calculate $B_n^k$, we employ the Stevens operator equivalent method \cite{chen2023magnetic}. $B_n^k$ can then be written as:
\begin{equation}
    B^{k}_{n} = -\dfrac{Q_{\rm M}}{4 \pi \epsilon_{0}} 
\sum_{l=1}^{L} \frac{4\pi}{(2n+1)} 
\frac{(-1)^{k} Y_{n}^{-k} (\theta_{l}, \phi_{l})}{R_{l}^{(n+1)}} q_{l} \langle r^{n} \rangle y^{k}_{n} \theta_{n}
    \tag{S4}
\end{equation}

where $Q_{\rm M}$ is the total charge of a Ho adatom,  $\varepsilon_0$ is the dielectric constant, $Y^{k}_{n}$ is the spherical harmonics, $R_{l}$ is the distance between the Ho adatom and the surrounding ligand atoms and $q_l$ is the charge on the ligand atom, the sum is over all $L$ ligand atoms. $\langle r^{n} \rangle$ can be derived from the references for the Ln ions \cite{freeman1962theoretical}. $y^{k}_{n}$ is the numerical coefficient occurring in $Y^{k}_{n}$, 
$\theta_{n}$ is the multiplicative factor \cite{stevens1952matrix}.  We used the crystal field coordination and the charge shown in Table 2 and Table 3 \cite{thesisTobias}. Table 4 summarizes the resulting coeffcients of the Stevens operators for Ho$_{\rm top}$ and Ho$_{\rm bridge}$ obtained from Eq. (S4). In Table 4, Ho$_{\rm top}$ exhibits a large uniaxial anisotropy term ($B^{0}_{2}$ = $-1387$ $\mu$eV), while Ho$_{\rm bridge}$ shows a large transverse anisotropy term ($B^{2}_{2}$ = $1323$ $\mu$eV).

\begin{table}[h]
\centering
\caption{Vertical distances ($d_{\perp}$) and lateral distances ($d_{\parallel}$) (in pm) between the Ho$_{\rm top}$ and the surrounding ions \cite{thesisTobias}.}
\begin{tabular}{l c c c}
\hline
Ion & Charge & $d_{\perp}$ (pm) & $d_{\parallel}$ (pm) \\
\hline
O (underneath) & $-2e$ & 213 & 0 \\
Mg             & $+2e$ & 276 & 213 \\
O              & $-2e$ & 275 & 294 \\
\hline
\end{tabular}
\label{tab:Ho_top}
\end{table}

\begin{table}[h]
\centering
\caption{Vertical distances ($d_{\perp}$) and lateral distances ($d_{\parallel}$) (in pm) between the Ho$_{\rm bridge}$ and the surrounding ions \cite{thesisTobias}.}
\begin{tabular}{l c c c}
\hline
Ion & Charge & $d_{\perp}$ (pm) & $d_{\parallel}$ (pm) \\
\hline
Mg             & $+2e$ & 248 & 157 \\
O              & $-2e$ & 173 & 139 \\
\hline
\end{tabular}
\label{tab:Ho_bridge}
\end{table}

\begin{table}[h]
\centering
\caption{Calculated crystal field parameters used in effective spin Hamiltonian for Ho$_{\rm top}$ and Ho$_{\rm bridge}$. Here we used $Q_{\rm M} = +3e$ \cite{natterer2017reading}.}
\label{tab:crystal_field_parameters}
\begin{tabular}{|c|c|c|}
\hline
 & Ho$_{\rm top}$ & Ho$_{\rm bridge}$ \\ \hline
$B^{0}_{2}$ & $-1387$ $\mu$eV & $-563$ $\mu$eV \\ \hline
$B^{0}_{4}$ & $-831$ neV & $359$ neV \\ \hline
$B^{0}_{6}$ & $-3.67$ neV & $1.56$ neV \\ \hline
$B^{2}_{2}$ &  & $1323$ $\mu$eV \\ \hline
$B^{2}_{4}$ &  & $2.49$ $\mu$eV \\ \hline
$B^{4}_{4}$ & $231$ neV & $-533$ neV \\ \hline
$B^{2}_{6}$ &  & $9.31$ neV \\ \hline
$B^{4}_{6}$ & $1.53$ neV & $11.5$ neV \\ \hline
$B^{6}_{6}$ &  & $1.20$ neV \\ \hline
\end{tabular}
\end{table}

 \vspace{25mm}

The Zeeman term in the Hamiltonian describes the interaction between the Ho spins 
and the external magnetic field. It reads:

\begin{equation}
H_{\rm Z} = g_{\rm{eff}}J_{\rm{z}} B \mu_{\rm{B}}
    \tag{S5}
\end{equation}
In this study, we consider the ground state with $J_z = 8$ and $g_{\rm eff} = 1.25$ \cite{forrester2019quantum}. Table 5 shows the eigenstates of the ground state and metastable state (Ho$\uparrow$ and Ho$\downarrow$) for Ho$_{\rm top}$ and Ho$_{\rm bridge}$ ($| \psi_0 \rangle$ and $| \psi_1 \rangle$ correspond to the eigenvectors of Ho$\uparrow$ and Ho$\downarrow$ in Fig. 1b and Fig. S3b, respectively), obtained by diagonalizing Eq. (S1) at $B = 3.0$ T.

The coefficient of transition probability between Ho$\uparrow$ and Ho$\downarrow$, induced by either electron spin–spin scattering or spin–phonon scattering, can be calculated using the following expression: \cite{ternes2015spin,thesisTobias,thesisOtte}:

\begin{align}
I_{01}
&= \tfrac{1}{2} \Bigl\{
   \left| \langle \psi_1 | {J}_+ | \psi_0 \rangle \right|^2
 + \left| \langle \psi_1 | {J}_- | \psi_0 \rangle \right|^2
 + 2 \left| \langle \psi_1 | {J}_z | \psi_0 \rangle \right|^2
 \Bigr\}
     \tag{S6}
\end{align} 
The first two terms enable $\Delta m = \pm 1$ transitions and the third term allows $\Delta m = 0$, where $\Delta m$ is the difference in the magnetization quantum number.

 Fig. 1b shows the corresponding energy-level diagram for Ho$_{\rm top}$ obtained by diagonalization of Eq.~(S1) at $B$ = 3.0 T. We obtain ground state and metastable state (Ho$\uparrow$ and Ho$\downarrow$ in Fig. 1b) with $\langle J_z \rangle = -8.0$ and $\langle J_z \rangle = +8.0$. As shown in Table~5, these states are composed of $\lvert m \rangle = \pm \lvert 8 \rangle$ states with more than \SI{99}{\percent} weight. No $\Delta m = \pm 1$ or $\Delta m = 0$ transitions are allowed between Ho$\uparrow$ and Ho$\downarrow$. Thus, according to Eq.~(S6), Ho$_{\rm top}$ exhibits a suppression of direct transition.

\begin{figure*}
\centering
\includegraphics[width=1.0\linewidth] {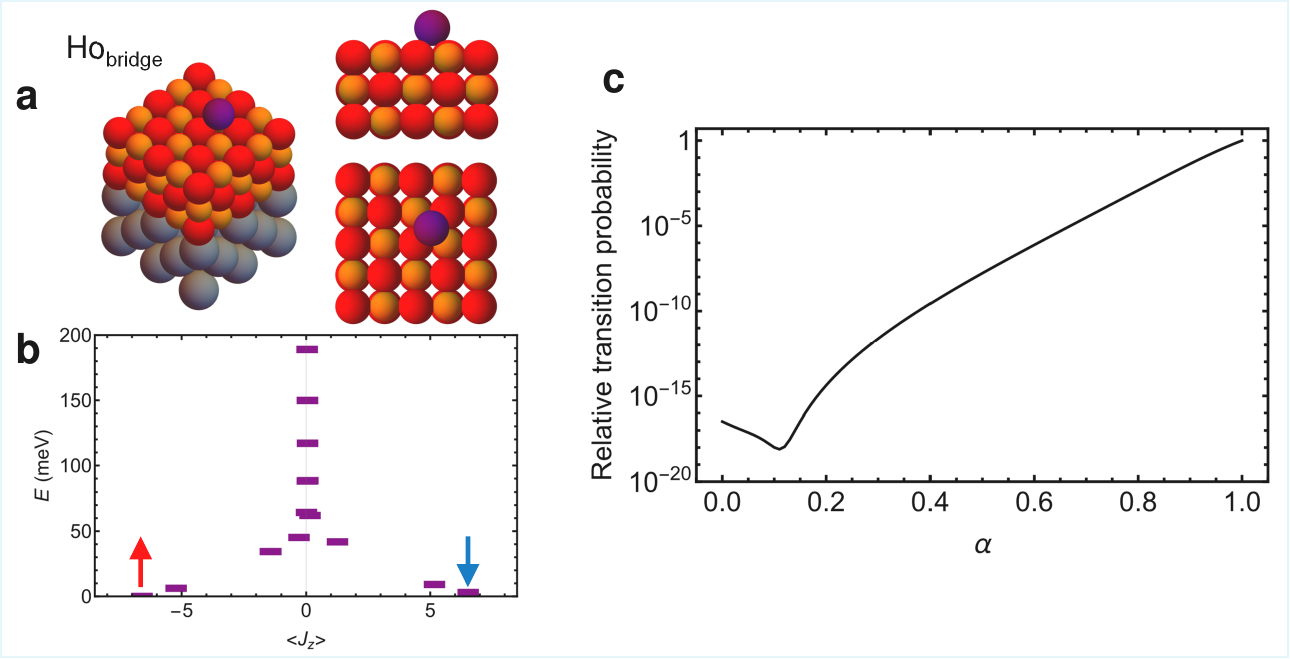}
\caption{\textbf{Energy diagram of Ho$_{\rm bridge}$.}
(a) Three-dimensional, top, and side views of the adsorption configuration of a Ho atom at the bridge site in the low-symmetry $C_{2v}$ position on MgO/Ag(100). Purple ball: Ho atom, orange ball: Mg atom, red ball: O atom, gray ball: Ag atom. (b) Calculated
eigenvalues of bridge-site Ho in low-symmetry $C_{2v}$ on MgO/Ag(100). Bridge-site Ho in low-symmetry $C_{2v}$ configuration shows strongly superposed angular momentum states even at large magnetic fields of $B =  \SI{3.0}{\tesla}$. The red and blue arrows in (b) indicate the ground state and metastable state (Ho↑ and Ho↓). In section 3 of the supporting information, the corresponding eigenvectors for Ho↑ and Ho↓ in (b) are denoted by $| \psi_0 \rangle$ and $| \psi_1 \rangle$, respectively. (c) A relative transition probability between the ground state and metastable state ($| \psi_0 \rangle$ and $| \psi_1 \rangle$) as a function of $\alpha$.}\label{fig:FigS3}
\end{figure*}

On the other hand, Fig.~\ref{fig:FigS3}b shows the energy-level diagram for Ho$_{\rm bridge}$ obtained by diagonalizing Eq.(S1) at $B = 3.0$ T. The ground state and metastable state (Ho$\uparrow$ and Ho$\downarrow$ in Fig.S3b) have $\langle J_z \rangle = -6.59$ and $\langle J_z \rangle = +6.53$. As indicated in Table~5, these states exhibit a $\Delta m = 0$ transition with an energy difference of 3.00 meV. 

The experimental results suggest that spin switching between Ho↑ and Ho↓ occurs when Ho$_{\rm top}$ moves toward Ho$_{\rm bridge}$ but does not fully reach it, due to the force exerted by the Co tip. To investigate this, we calculate the relative transition probability between the Ho$\uparrow$ and Ho$\downarrow$ as a function of a linear combination of $H_{\mathrm{cf}\, C_{2v}}$ and $H_{\mathrm{cf}\, C_{4v}}$ \cite{natterer2018thermal}. Specifically, we first diagonalize the Hamiltonian $H = \alpha H_{\mathrm{cf}\, C_{2v}} + (1-\alpha) H_{\mathrm{cf}\, C_{4v}} + H_{\rm Z}$, and then evaluate the relative coefficient of transition probability using Eq.~(S6) as a function of $\alpha$. As shown in Fig.~\ref{fig:FigS3}c, $H_{\mathrm{cf}\, C_{4v}}$ ($\alpha=0$) exhibits a lower transition probability than $H_{\mathrm{cf}\, C_{2v}}$ ($\alpha=1$), as expected. Furthermore, Fig.~\ref{fig:FigS3}c shows that the relative transition probability changes approximately exponentially with respect to $\alpha$.

\begin{table}[h]
\centering
\caption{Calculated eigenvectors for the ground state and metastable state, obtained by diagonalization of (S1) at $B$ = 3.0 T.}
\label{tab:eigenvectors_S8}
\begin{tabular}{lcccc}
\toprule
 & \multicolumn{2}{c}{Ho$_{\rm top}$} & \multicolumn{2}{c}{Ho$_{\rm bridge}$} \\
$\lvert m \rangle$ & $\lvert \psi_0 \rangle$ & $\lvert \psi_1 \rangle$ & $\lvert \psi_0 \rangle$ & $\lvert \psi_1 \rangle$ \\
\midrule
$\lvert +8 \rangle$ & 0.00 & 0.9999 & 0.073 & 0.72 \\
$\lvert +7 \rangle$ & 0.00 & 0.00 & 0.00 & 0.00 \\
$\lvert +6 \rangle$ & 0.00 & 0.00 & -0.067 & -0.59 \\
$\lvert +5 \rangle$ & 0.00 & 0.00 & 0.00 & 0.00 \\
$\lvert +4 \rangle$ & 0.00 & -0.0017 & 0.046 & 0.32 \\
$\lvert +3 \rangle$ & 0.00 & 0.00 & 0.00 & 0.00 \\
$\lvert +2 \rangle$ & 0.00 & 0.00 & -0.042 & -0.15 \\
$\lvert +1 \rangle$ & 0.00 & 0.00 & 0.00 & 0.00 \\
$\lvert 0  \rangle$ & 0.00 & 0.00 & 0.067 & 0.058 \\
$\lvert -1 \rangle$ & 0.00 & 0.00 & 0.00 & 0.00 \\
$\lvert -2 \rangle$ & 0.00 & 0.00 & -0.14 & -0.012 \\
$\lvert -3 \rangle$ & 0.00 & 0.00 & 0.00 & 0.00 \\
$\lvert -4 \rangle$ & -0.0017 & 0.00 & 0.31 & -0.024 \\
$\lvert -5 \rangle$ & 0.00 & 0.00 & 0.00 & 0.00 \\
$\lvert -6 \rangle$ & 0.00 & 0.00 & -0.58 & 0.067\\
$\lvert -7 \rangle$ & 0.00 & 0.00 & 0.00 & 0.00 \\
$\lvert -8 \rangle$ & 0.9999 & 0.00 & 0.73 & -0.098 \\
\midrule
Energy (meV) & 0.00 & 3.47 & 0.00 & 3.00 \\
\bottomrule
\end{tabular}
\end{table}

\vspace{120mm}

\section{Fabrication of Co tip}
\begin{figure*}
\centering
\includegraphics[width=1.0 \linewidth] {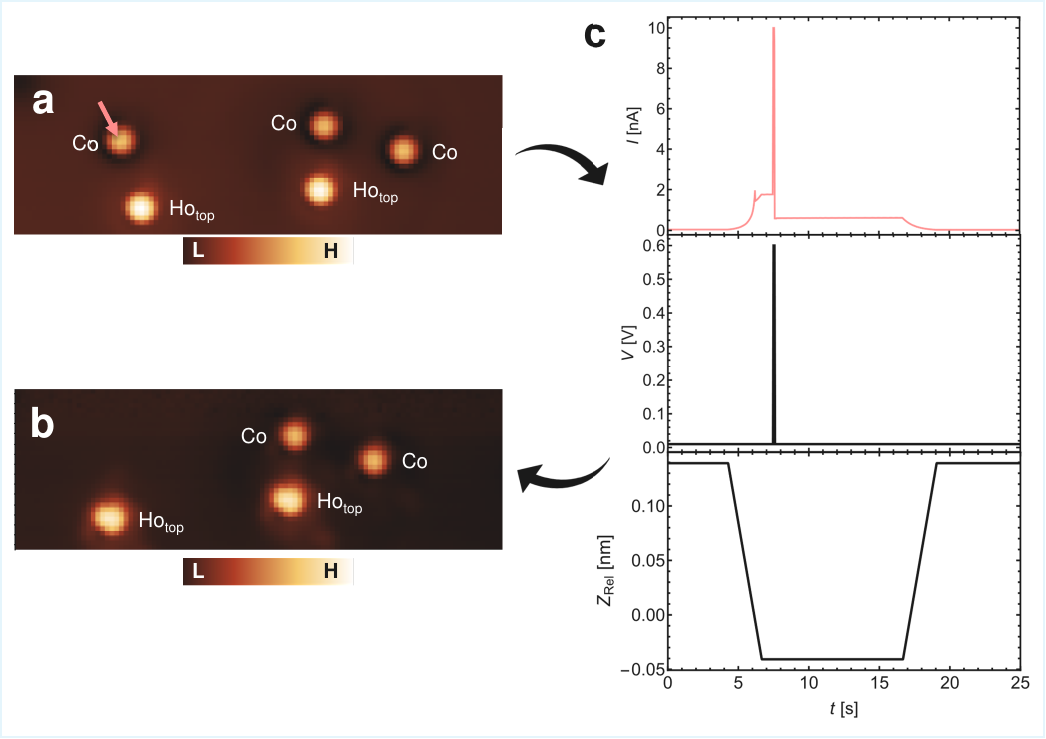}
\caption{\textbf{Picking up a Co adatom from MgO/Ag(001).} (a, b) Consecutive STM topography images of coadsorbed Ho$_{\rm top}$ and Co on MgO/Ag(001) before and after picking up Co from the surface. Imaging conditions: constant-current mode, \textit{$I$} = \SI{1.0}{\nA}, \textit{$V$} = \SI{200}{\mV}, \textit{$B$} = \SI{3.0}{\tesla}. (c) Time dependence of the tunneling current, bias voltage, and tip--sample distance during the pickup of a Co adatom from the surface.
}\label{fig:FigS4}
\end{figure*}

In our experiment, the fabrication of the magnetic tip was carried out as follows \cite{thesisBaumann}. First, we scanned the Co adatoms using a non-magnetic  (W or Ag) tip. Fig.~\ref{fig:FigS4}a shows the STM topography image of coadsorbed Co and Ho$_{\rm{top}}$ adatoms on MgO obtained by the non-magnetic tip. Next, the non-magnetic tip was positioned above the target Co adatom (highlighted by the pink arrow in Fig.~\ref{fig:FigS4}a) and moved vertically toward the Co adatom by approximately \SI{0.18}{\nano\metre} from the STM set point (Fig.~\ref{fig:FigS4}c). After that, a bias pulse of 600 mV for 0.1 s was applied (Fig.~\ref{fig:FigS4}c). Subsequently, we rescanned the same area and confirmed that the target Co adatom had disappeared, suggesting that Co atoms were picked up onto the tip apex (Fig.~\ref{fig:FigS4}b). The Ho$_{\rm{top}}$ was then probed to observe spin switching using constant current mode ($I = 1$ nA and $V = 200$ mV), verifying the formation of the stable Co tip (see also Fig.~\ref{fig:FigS5}b and Fig.~\ref{fig:FigS5}e). This procedure was typically repeated several times to ensure a stable Co tip.

\section{Probing Ho and Co atoms on MgO/Ag(001) using non-magnetic tip and magnetic tip}
\begin{figure*}
\centering
\includegraphics[width=0.9 \linewidth] {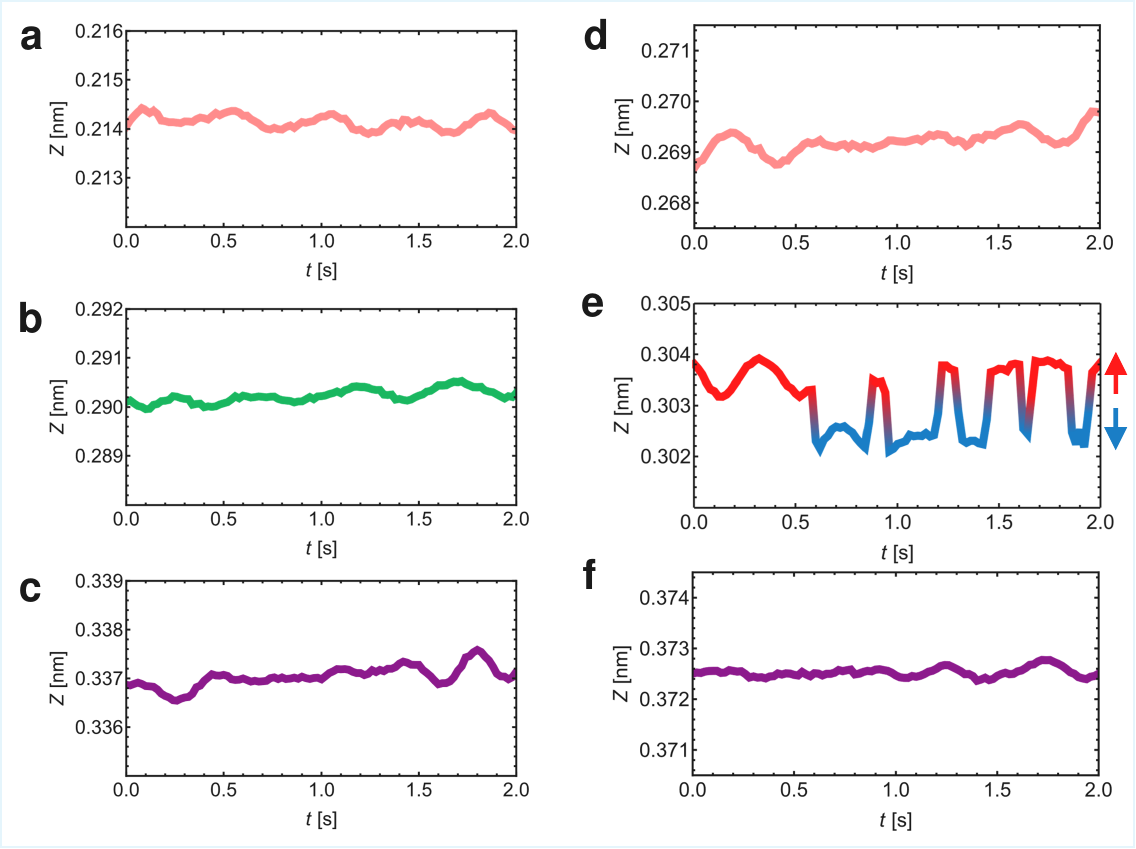}
\caption{\textbf{Probing Co, Ho$_{\rm{top}}$, and Ho$_{\rm{bridge}}$ using non-magnetic tip and magnetic tip.}
(a--c) Time dependence of the tip–sample distance measured on top of Co, Ho$_{\rm{top}}$, and Ho$_{\rm{bridge}}$ using a non-magnetic tip. (d--f) Time dependence of the tip–sample distance measured on the same Co, Ho$_{\rm{top}}$, and Ho$_{\rm{bridge}}$ as in (a--c), immediately after picking up a Co adatom from the surface with the tip. The red and blue in (e) indicate Ho$\uparrow$ and Ho$\downarrow$. Measurement conditions: constant-current mode, \textit{$I$} = \SI{1.0}{\nA}, \textit{$V$} = \SI{200}{\mV},  \textit{$B$} = \SI{3.0}{\tesla}.}\label{fig:FigS5}
\end{figure*}

Fig.~\ref{fig:FigS5}(a--c) shows the time dependence of the tip–sample distance recorded over Ho$_{\rm top}$, Ho$_{\rm bridge}$, and Co adatoms on MgO in constant-current mode using a non-magnetic tip, respectively. Fig.~\ref{fig:FigS5}(d--f) presents the corresponding measurements on the same Ho$_{\rm top}$, Ho$_{\rm bridge}$, and Co adatoms as in Figs.~\ref{fig:FigS5}(a--c), immediately after a Co adatom was picked up from the surface, as described in Fig.~\ref{fig:FigS4}. In Figs.~\ref{fig:FigS5}(a--f), the tunneling current and the bias voltage were set to values suitable for inducing spin switching by current ($I = 1$ nA and $V = 200$ mV) \cite{natterer2017reading,natterer2018thermal}. The difference in tunneling magnetoresistance between the Ho$\uparrow$ and Ho$\downarrow$ states leads to a change in the tip–sample distance; consequently, spin switching appears as a random telegraph signal \cite{natterer2017reading,natterer2018thermal}. As shown in Fig.~\ref{fig:FigS5}e, the spin switching is observed for Ho$_{\rm{top}}$ when using a magnetic tip.

\section{Tunneling current simultaneously recorded with $\Delta f$ in Fig.2a}
\begin{figure*}
\centering
\includegraphics[width=0.6 \linewidth] {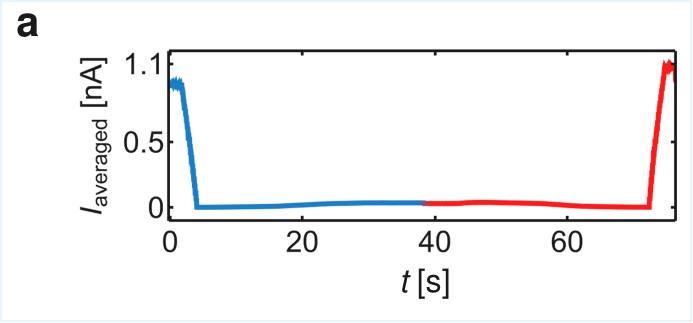}
\caption{\textbf{Averaged tunneling current simultaneously obtained with $\Delta$\textit{$f$} in Fig.2a.}
(a) Time evolution of averaged tunneling current, simultaneously recorded with $\Delta$\textit{$f$} in Fig.2a. The blue and red correspond to the Ho$\downarrow$ and Ho$\uparrow$.}
\label{fig:FigS6}
\end{figure*}

Fig.~\ref{fig:FigS6}a shows the averaged tunneling current recorded simultaneously with $\Delta f$ in Fig. 2a. In Fig.~\ref{fig:FigS6}a, the averaged tunneling current after restoring the bias voltage ($t=76$ s) is larger than that before reducing the bias voltage ($t=0$ s). These results further support the spin switching from Ho↓ to Ho↑ by approaching the tip toward Ho$_{\rm{top}}$.

\section{Conductance and force spectroscopy measured on top of the Ho$\uparrow$ and the Ho$\downarrow$}
\begin{figure*}
\centering
\includegraphics[width=1.00 \linewidth] {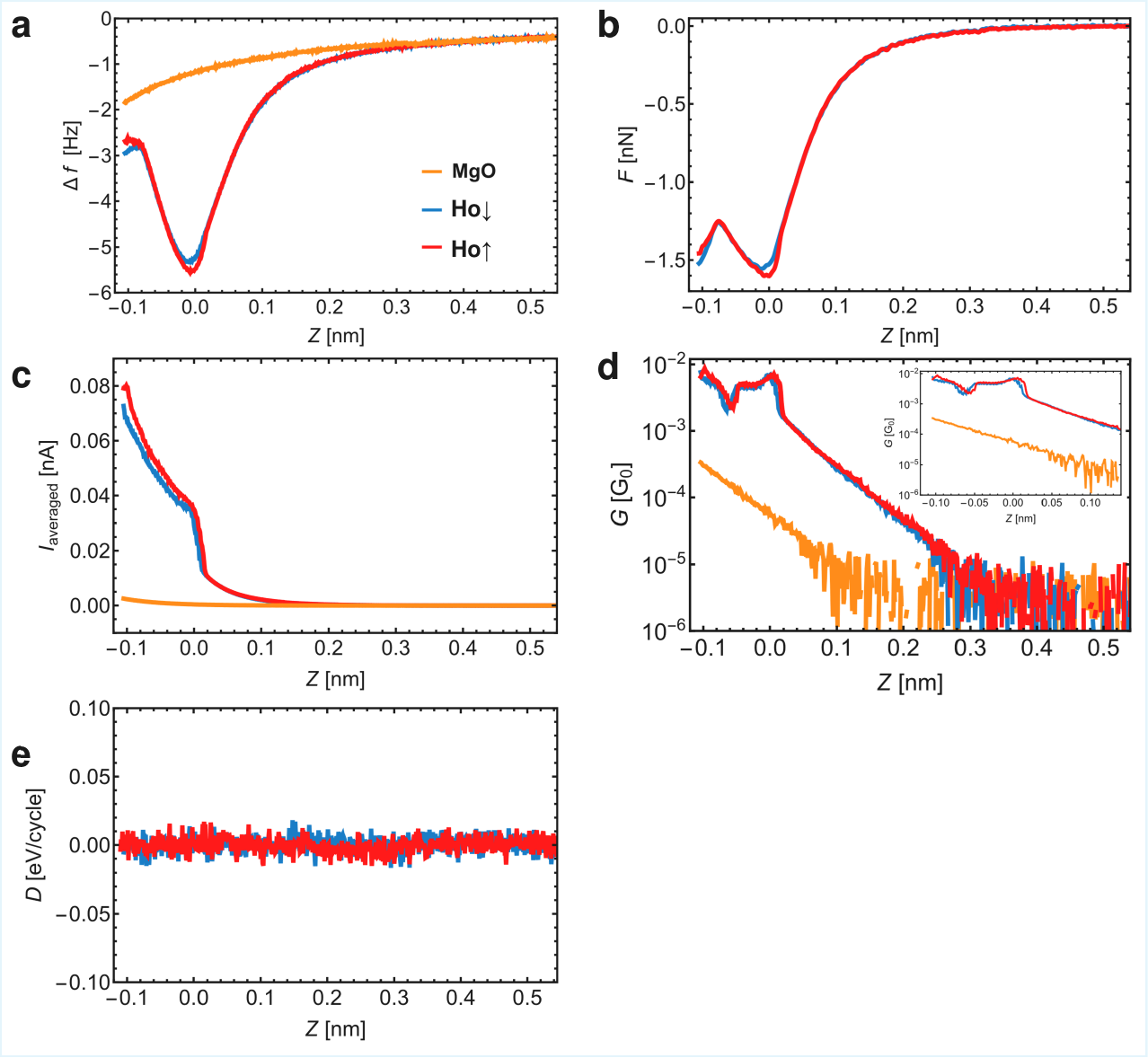}
\caption{\textbf{Full data sets for Fig.3.}
(a) The distance dependence of frequency shift obtained on top of the Ho$\uparrow$ ($\Delta$\textit{$f$}$_{\rm{Ho\uparrow}}$(\textit{$z$}), red solid curve), Ho$\downarrow$ ($\Delta$\textit{$f$}$_{\rm{Ho\downarrow}}$(\textit{$z$}), blue solid curve) and MgO surface ($\Delta f_{\mathrm{MgO}}(z)$, orange solid curve). Measurement conditions: $V = \SI{+200}{\micro\volt}$ and $B = \SI{+3.0}{\tesla}$. 
 (b) Short-range force obtained on top of Ho$\uparrow$ (\textit{$F$}$_{\rm{Ho\uparrow}}$(\textit{$z$}), red) and Ho$\downarrow$ (\textit{$F$}$_{\rm{Ho\downarrow}}$(\textit{$z$}), blue).  
\textit{$F$}$_{\rm{Ho\uparrow}}$(\textit{$z$}) and \textit{$F$}$_{\rm{Ho\downarrow}}$(\textit{$z$}) were calculated from $\Delta$\textit{$f$}$_{\rm{Ho\uparrow SR}}$(\textit{$z$}) = $\Delta$\textit{$f$}$_{\rm{Ho\uparrow}}$(\textit{$z$}) $-$ $\Delta$\textit{$f$}$_{\rm{MgO}}$(\textit{$z$}) and $\Delta$\textit{$f$}$_{\rm{Ho\downarrow SR}}$(\textit{$z$}) = $\Delta$\textit{$f$}$_{\rm{Ho\downarrow}}$(\textit{$z$}) $-$ $\Delta$\textit{$f$}$_{\rm{MgO}}$(\textit{$z$}) \cite{sader2004accurate}. (c) The distance dependence of the averaged tunneling current simultaneously recorded on top of Ho$\uparrow$ (red solid curve), Ho$\downarrow$ (blue solid curve) and MgO surface (orange solid curve) with $\Delta$\textit{$f$} in (a). (d) The distance dependence of conductance at the lowest turnaround point of the tip oscillation cycle, obtained by deconvolving the time-averaged tunneling current in (c) \cite{sader2010accurate}. Ho$\uparrow$ (red solid curve), Ho$\downarrow$ (blue solid curve) and MgO surface (orange solid curve) in (d). Inset shows the enlarged image of the range $-0.11$ nm $\leq z \leq 0.14$ nm. (e) Distance dependence of the dissipation simultaneously recorded with $\Delta$\textit{$f$}$_{\rm{Ho\downarrow}}$($z$) and $\Delta$\textit{$f$}$_{\rm{Ho\uparrow}}$($z$) in (a). Ho$\uparrow$ (red solid curve) and Ho$\downarrow$ (blue solid curve) in (e).}\label{fig:FigS7}
\end{figure*}

Fig.~\ref{fig:FigS7}a shows the distance dependence of $\Delta$\textit{$f$} recorded on top of the Ho$\uparrow$, Ho$\downarrow$ and MgO substrate, respectively. Note that Fig.~3a shows the range $-0.11$ nm $\leq$ $z$ $\leq$ $0.14$ nm of Fig.~\ref{fig:FigS7}a. The $\Delta$\textit{$f$}($z$) recorded over a Ho adatom includes a long-range force from the MgO substrate. Therefore, $\Delta$\textit{$f$}($z$) measured on top of the MgO surface was subtracted from that measured over the Ho adatom to eliminate its background component. Fig.~\ref{fig:FigS7}b shows the distance dependence of the background-subtracted short-range force. Fig.~3b shows the range $-0.11$ nm $\leq$ $z$ $\leq$ $0.14$ nm of Fig.~\ref{fig:FigS7}b.

Fig.~\ref{fig:FigS7}c shows the distance dependence of the time-averaged tunneling current, recorded simultaneously with $\Delta$\textit{$f$} in Fig.~\ref{fig:FigS7}a. Fig.~\ref{fig:FigS7}d shows the distance dependence of conductance at the lowest turnaround point of the tip oscillation cycle, obtained by deconvolving the time-averaged tunneling current in Fig.~\ref{fig:FigS7}c. 

In Fig.~\ref{fig:FigS7}b and Fig.~\ref{fig:FigS7}d, we observe the short-range force of \textit{$F$}$_{\rm{Ho\uparrow}}$($z = 0.00$ nm) = $-1.60$ nN and \textit{$F$}$_{\rm{Ho\downarrow}}$($z = 0.00$ nm) = $-1.55$ nN, concomitant with a point-contact conductance \cite{ternes2011interplay,kroger2008contact}. For smaller tip–sample distances, the force increases up to \textit{$F$}$_{\rm{Ho\uparrow}}$($z = -0.08$ nm) = –1.25 nN and \textit{$F$}$_{\rm{Ho\downarrow}}$($z = -0.08$ nm) = –1.25 nN, and then decreases to \textit{$F$}$_{\rm{Ho\uparrow}}$($z = -0.10$ nm) = –1.40 nN and \textit{$F$}$_{\rm{Ho\downarrow}}$($z = -0.10$ nm) = –1.50 nN, as discussed in Fig.~3.

Fig.~\ref{fig:FigS7}e shows the dissipation signal as a function of tip--sample distance, simultaneously recorded with $\Delta$\textit{$f$}$_{\rm{Ho\uparrow}}$($z$) and  $\Delta$\textit{$f$}$_{\rm{Ho\downarrow}}$($z$) in Fig.~\ref{fig:FigS7}a. The negligible dissipation signal indicates that the position of the Ho adatom is determined for each tip--sample distance and that it is independent of the tip oscillation. In other words, when the Ho adatom moves laterally, the force between the Ho adatom and the Co tip exhibits no hysteresis throughout the tip oscillation.

\section{$\Delta$\textit{$f$}($z$) measured on top of the Ho$_{\rm{top}}$}
\begin{figure*}
\centering
\includegraphics[width=0.8 \linewidth] {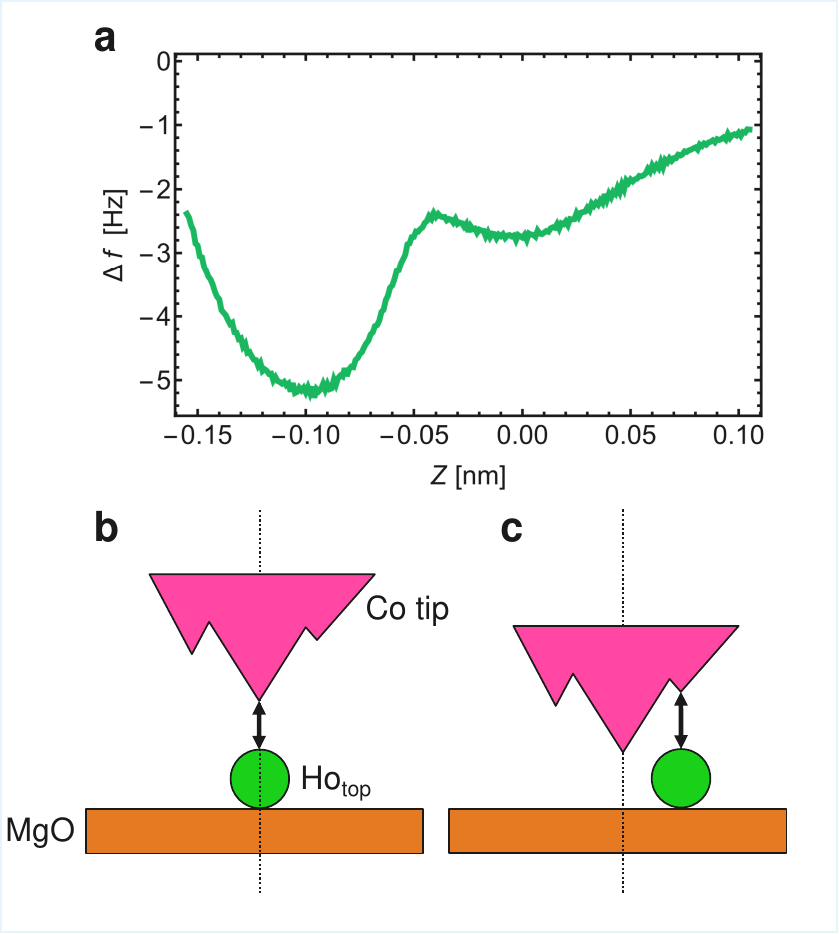}
\caption{\textbf{$\Delta$\textit{$f$}($z$) measured on top of the Ho$_{\rm{top}}$}.
(a) The distance dependence of frequency shift obtained on top of the Ho adatom. (b,c) Schematic illustration of the lateral motion of the Ho adtom by approaching the tip. Large tip sample distance (b) and small tip sample distance (c). Measurement conditions: $V = \SI{+200}{\micro\volt}$ and $B = \SI{+3.0}{\tesla}$.}\label{fig:FigS8}
\end{figure*}

Fig.~\ref{fig:FigS8}a shows the distance dependence of $\Delta f$ recorded on top of the Ho$_{\rm{top}}$. As the Co tip approaches the Ho adatom, $\Delta f$ reaches a local minimum of $\Delta f$ = $-2.7$ Hz near $z = 0$ nm. Further reduction of the distance $z$ first leads to a slight increase and then decreases to another local minimum of $\Delta f = -5.1$ Hz at $z = -0.10$ nm. 

This trend is also observed in Fig.~\ref{fig:FigS7}a, Fig.~\ref{fig:FigS11}b and Fig.~\ref{fig:FigS12}a and can be explained by the lateral motion of the Ho adatom beneath the tip. Fig.~\ref{fig:FigS8}b and Fig.~\ref{fig:FigS8}c provide a schematic illustration of this lateral motion. When the tip–sample distance is large (Fig.~\ref{fig:FigS8}b), the $\Delta f$ curve near $z = 0~\text{nm}$ reflects the interaction between the frontmost Co atom of the tip and the Ho adatom. As the tip–sample distance decreases (Fig.~\ref{fig:FigS8}c), the Ho adatom laterally moves beneath the tip and begins to interact with Co atoms on the tip other than the frontmost atom. Similar behavior has been discussed for the tip apex atom on NaCl(100), and for the Si(001) \cite{schirmeisen2006single, bamidele2012complex}. Hence, Fig.~\ref{fig:FigS8} supports that Ho$_{\rm top}$ moves laterally on MgO, due to the force exerted by the Co tip.

\section{Controlling Ho spin from Ho↑ to Ho↓ by approaching the tip}
\begin{figure*}
\centering
\includegraphics[width=0.8 \linewidth] {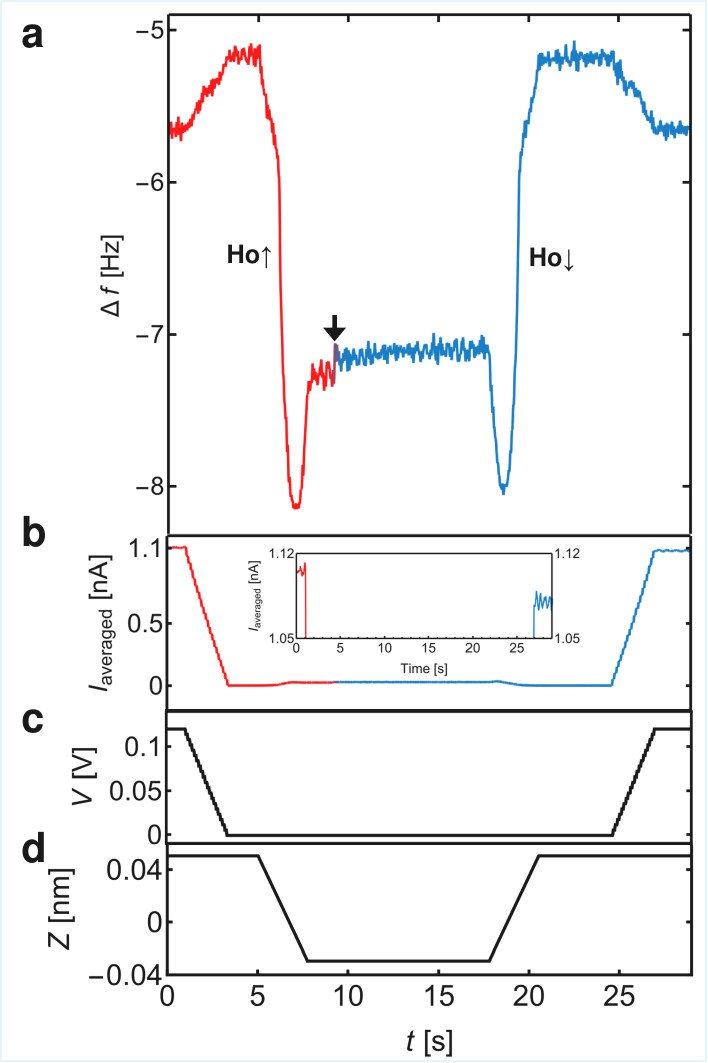}
\caption{\textbf{Controlling Ho spin from Ho↑ to Ho↓ by approaching the tip.} (a--d) $\Delta f(t)$ spectra (a) and averaged tunneling current (b) measured on top of Ho$_{\rm{top}}$ while varying the bias voltage (c) and the tip sample distances (d). The red and blue in (a,b) indicate the Ho$\uparrow$ and Ho$\downarrow$. At $8~\mathrm{s} \leq t \leq 18~\mathrm{s}$, the transition from the Ho$\uparrow$ to Ho$\downarrow$ state can be detected by a sudden jump in $\Delta f(t)$, marked by the black arrow. The inset in (b) shows the enlarged image of the averaged tunneling current at $t = 0$~s and $t = 29$~s. Measurement conditions: \textit{$B$}  = \SI{+ 3.0}{\tesla}. }
%MgO = 2 ML}
\label{fig:FigS9}
\end{figure*}

In Fig.~\ref{fig:FigS9}, employing the same approach of bringing the tip toward Ho$_{\rm{top}}$ as in Fig. 2a, we show the reverse process, namely the readout of Ho↑ and Ho↓, as well as spin switching from Ho↑ to Ho↓. Fig.~\ref{fig:FigS9}a and Fig.~\ref{fig:FigS9}b show typical $\Delta$\textit{$f$} and averaged tunneling current as a function of time, measured on top of Ho$_{\rm{top}}$ while varying the bias voltage (Fig.~\ref{fig:FigS9}c) and the tip--sample distances (Fig.~\ref{fig:FigS9}d). As discussed in Fig.2, first, the lateral position of the tip was fixed above Ho$_{\rm{top}}$, and its spin state was stabilized in the desired configuration (in this case, Ho↑) by lowering the bias voltage from \textit{V} = \SI{+120}{mV} to \textit{V} = \SI{+200}{\micro\volt}, well below the threshold for current-induced spin switching (\SI{0}{\second} $\leq t \leq$ \SI{3}{\second}).  Once set, the Ho↑ state was measured via $\Delta f$ during the tip approach (\SI{5}{\second} $\leq t \leq$ \SI{8}{\second}). To switch the spin state from Ho↑ to Ho↓, the tip was brought to a specific distance (\textit{z} = $-0.03$ nm), exceeding the threshold required to induce spin switching.
The spin state of the Ho adatom was then probed at this distance with a fixed probe time of 10 s (\SI{8}{\second} $\leq t \leq$ \SI{18}{\second}), and the transition from Ho↑ to Ho↓ was detected as a sudden jump in $\Delta f$, indicated by the black arrow in Fig.~\ref{fig:FigS9}a. Afterward, the Ho↓ was measured from $\Delta f$ by retracting the tip to its original tip sample distance (\SI{18}{\second} $\leq t \leq$ \SI{21}{\second}).  Finally, the bias voltage was restored from \textit{V} = \SI{+200}{\micro\volt} to its original value of \textit{V} = \SI{+120}{mV} (\SI{25}{\second} $\leq t \leq$ \SI{29}{\second}). 

In Fig.~\ref{fig:FigS9}a, the minimum of $\Delta f$ obtained on top of Ho↑ is smaller than that for Ho↓, indicating that the spin state of Ho can be read from $\Delta f$. Hence, Ho↑ can be switched to Ho↓ by approaching the tip. Moreover, in Fig.~\ref{fig:FigS9}b, the averaged tunneling current after restoring the bias voltage ($t=29$ s) is smaller than that before reducing the bias voltage ($t=0$ s). These observations further support the spin switching from Ho↑ to Ho↓ by approaching the tip toward Ho$_{\rm{top}}$.

\section{Ho spin switching induced by tunneling current}
Fig. S10(a–c) shows the time dependence of the tunneling current recorded over Ho$_{\rm{top}}$ at different tip–sample distances. The switching rates for various tip–sample distances are summarized in Fig. 4c.

\section{Atom manipulation from Ho$_{\rm top}$ to Ho$_{\rm bridge}$ by tip approach}
Here, we demonstrate that approaching the tip toward Ho$_{\rm top}$ induces atomic manipulation toward Ho$_{\rm bridge}$. Fig.~\ref{fig:FigS11}a shows an STM topography image of three Ho$_{\rm top}$ atoms obtained in constant-current mode at \textit{V} = \SI{+200}{mV}, confirming their initial adsorption sites. Among them, one Ho$_{\rm top}$ atom was selected as the target (gray arrow in Fig.~\ref{fig:FigS11}a). For this target Ho$_{\rm top}$, the tip was positioned above the center of the atom, and its spin state was stabilized in the desired configuration (here, Ho↓) by lowering the bias voltage from \textit{V} = \SI{+200}{mV} to \textit{V} = \SI{+200}{\micro\volt}, well below the threshold for current-induced spin switching. Subsequently, the tip was approached vertically by decreasing the tip height until spin switching was observed (see Fig.~\ref{fig:FigS11}b). The switching from Ho↓ to Ho↑ is evidenced by a sudden jump of about 0.2 Hz in $\Delta f$ at $t = 19$ s (see Fig.~\ref{fig:FigS11}c).

Next, after further approaching the tip toward Ho$_{\rm top}$ from the distance where spin switching occurred, the tip was held at a constant height ($23$ s $\leq t \leq 32$ s). As a result, a sudden jump of approximately 0.8 Hz in $\Delta f$, corresponding to the atom manipulation from Ho$_{\rm top}$ to Ho$_{\rm bridge}$, was observed at $t = 28$ s (see Fig.~\ref{fig:FigS11}c). Fig.~\ref{fig:FigS11}d shows an STM image of the same scan area as Fig.~\ref{fig:FigS11}a, recorded immediately after Fig.~\ref{fig:FigS11}b in constant-current mode at \textit{V} = \SI{+200}{mV}, revealing that the target Ho$_{\rm top}$ has transformed into a brighter spot corresponding to Ho$_{\rm bridge}$. These results demonstrates that approaching the tip to the center of Ho$_{\rm top}$ can induce a lateral displacement toward the bridge site, resulting in the formation of Ho$_{\rm bridge}$. Note that once the Ho atom relocates to the Ho$_{\rm bridge}$ position, it rarely returns to Ho$_{\rm top}$.

\section{Imaging Ho adatom at small tip--sample distances}
Fig.~\ref{fig:FigS12}a and Fig.~\ref{fig:FigS12}b show the $\Delta f$ and conductance recorded on top of Ho$\uparrow$ and Ho$\downarrow$ as a function of tip–sample distance.  To image Ho$\uparrow$ and Ho$\downarrow$ at small tip–sample distances, the tip height was set to $z = 0$ nm near the local minimum of $\Delta f$. Fig.~\ref{fig:FigS12}c and Fig.~\ref{fig:FigS12}d present $\Delta f$ images of Ho$\uparrow$ and Ho$\downarrow$ acquired at $z = 0$ nm. In Fig.~\ref{fig:FigS12}c, three distinct local minimum are observed on the Ho adatom (white dotted circle in Fig.~\ref{fig:FigS12}c). As discussed in previous results, these three features indicate a three-atom tip \cite{gretz2020identifying}. Due to the asymmetric shape of the tip apex, atomic relaxation from Ho$_{\rm top}$ to Ho$_{\rm bridge}$ can be induced even when the tip is vertically approached toward the center of Ho$_{\rm top}$ (the center of Ho$_{\rm top}$ refers to the peak position of Ho$_{\rm top}$ in the STM image). Notably, spin switching from Ho$\downarrow$ to Ho$\uparrow$ was observed at $z = -0.11$ nm in Fig.~\ref{fig:FigS12}a. 

\section{Force-based spin switching by different tip}
To investigate the effect of tip shape on spin switching, we show the spin switching distance by approaching the Co tip at the center of Ho$_{\rm top}$ using different tip shapes. We prepared different tips using the method shown in Fig.~\ref{fig:FigS4} and examined the spin switching distance for each tip, corresponding to a fixed switching rate (here, 0.1 Hz). As shown in Fig.~\ref{fig:FigS13}a, the spin switching distance varies significantly with tip shape. We further found that spin switching always occurs at distances shorter than the point-contact distance ($z < 0$). Therefore, we propose that atomic relaxation from Ho$_{\rm top}$ to Ho$_{\rm bridge}$ begins after the tip reaches point contact, and that spin switching occurs when the Ho atom moves laterally from the top site to some extent without fully reaching the bridge site. Fig.~\ref{fig:FigS13}a confirms the reproducibility of the force-based spin switching.

\section{Proposed force-based spin switching mechanism}
Here, we discuss two possible mechanisms of spin switching induced by force. The first possible mechanism involves spin–phonon coupling between Ho and the MgO substrate \cite{thesisTobias}. The Ho adatom on the MgO surface has a phonon density of states at approximately 4.7~meV \cite{donati2020unconventional}. Therefore, we propose that the relaxation from Ho$_{\rm top}$ to Ho$_{\rm bridge}$ allows spin switching, which can be induced by spin–phonon scattering. The second possible mechanism involves electron spin–spin scattering due to the proximity of the Co tip to the Ho atom \cite{ternes2015spin}. The thermal energy of conduction electrons is estimated as $k_{\mathrm{B}} T \approx 0.388$~meV at 4.5~K. Thus, the relaxation from Ho$_{\rm top}$ to Ho$_{\rm bridge}$ allows spin switching, which can be induced by electron spin-spin scattering.

When the asymmetric Co tip is in contact with Ho$_{\rm top}$, we cannot exclude the possibility that the asymmetric Co tip may also reduce the symmetry of Ho$_{\rm top}$. However, because the Co tip is metallic and electrically neutral, whereas the oxygen and magnesium atoms in MgO are negatively and positively charged, the reduction of symmetry of Ho$_{\rm top}$ due to the crystal field is expected to be larger than that caused by the asymmetric tip.



\clearpage




\clearpage

\clearpage

\clearpage

\clearpage

\clearpage

\clearpage

\clearpage

\clearpage

\clearpage

\clearpage

\begin{figure*}
\centering
\includegraphics[width=0.6 \linewidth] {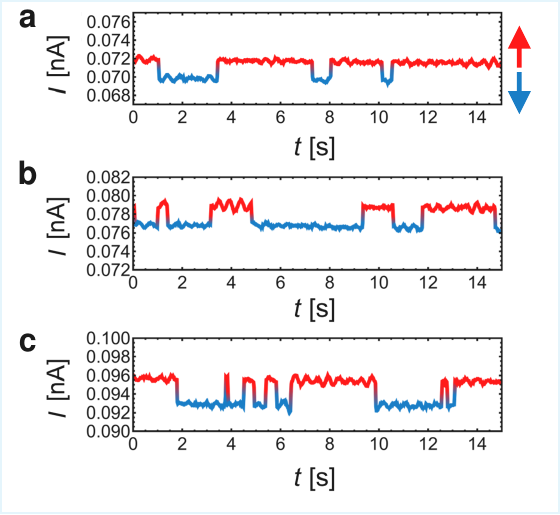}
\caption{\textbf{Spin switching induced by tunneling current.}
(a--c) Time dependence of the tunneling current measured on top of Ho$_{\rm{top}}$ while varying the tip–sample distance. The red and blue indicate the Ho$\uparrow$ and Ho$\downarrow$. Measurement conditions: constant-height mode, \textit{$V$} = \SI{150}{\mV}, \textit{$B$} = \SI{3.0}{\tesla}, (a) \textit{$z$} = \SI{292}{\pm}; (b) \textit{$z$} = \SI{287}{\pm}; (c) \textit{$z$} = \SI{277}{\pm}.}
\label{fig:FigS10}
\end{figure*}

\begin{figure*}
\centering
\includegraphics[width=1.00 \linewidth] {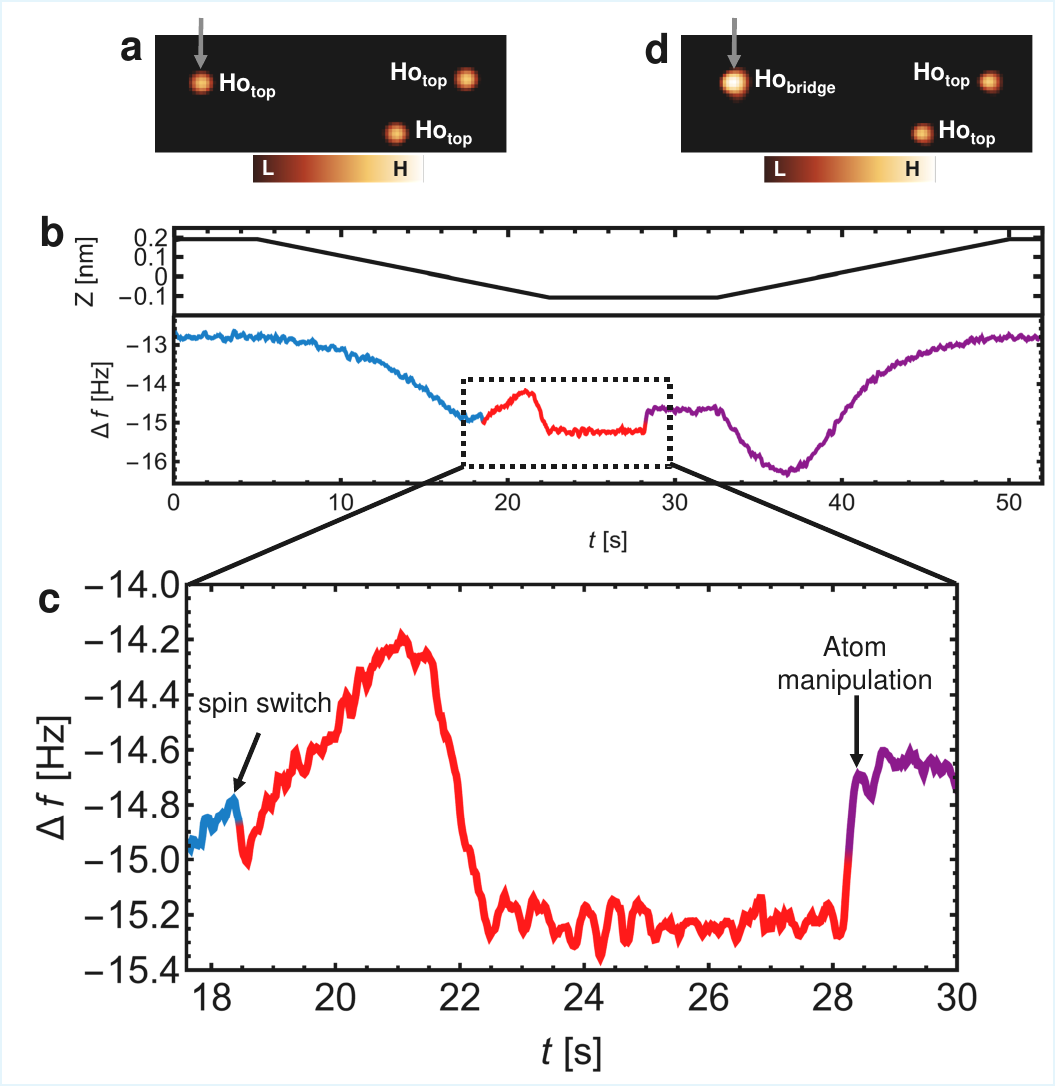}
\caption{\textbf{Atom manipulation induced by approaching the tip.} 
(a) STM topography image of Ho$_{\rm{top}}$ on MgO/Ag(001). The gray arrow indicates the target Ho$_{\rm{top}}$. Imaging conditions: constant-current mode, $I = \SI{1.0}{nA}$, $V = \SI{+200}{\milli\volt}$ and $B = \SI{+3.0}{\tesla}$.
(b) Time dependence of $\Delta f$ as a function of tip–sample distance. The blue, red and purple indicate the  Ho$\downarrow$, Ho$\uparrow$ and Ho$_{\rm{bridge}}$. Measurement conditions: $V = \SI{+200}{\micro\volt}$ and $B = \SI{+3.0}{\tesla}$.
(c) Enlarged view of (b) for $17~\mathrm{s} \leq t \leq 30~\mathrm{s}$, showing a sudden change in $\Delta f$ corresponding to spin switching and atom manipulation from Ho$_{\rm{top}}$ to Ho$_{\rm{bridge}}$. 
(d) STM topography image of the same area as in (a), acquired immediately after (b). Imaging conditions: constant-current mode, $I = \SI{1.0}{nA}$, $V = \SI{+200}{\milli\volt}$ and $B = \SI{+3.0}{\tesla}$.}\label{fig:FigS11}
\end{figure*}

\clearpage 
\clearpage
\begin{figure*}
\centering
\includegraphics[width=1 \linewidth] {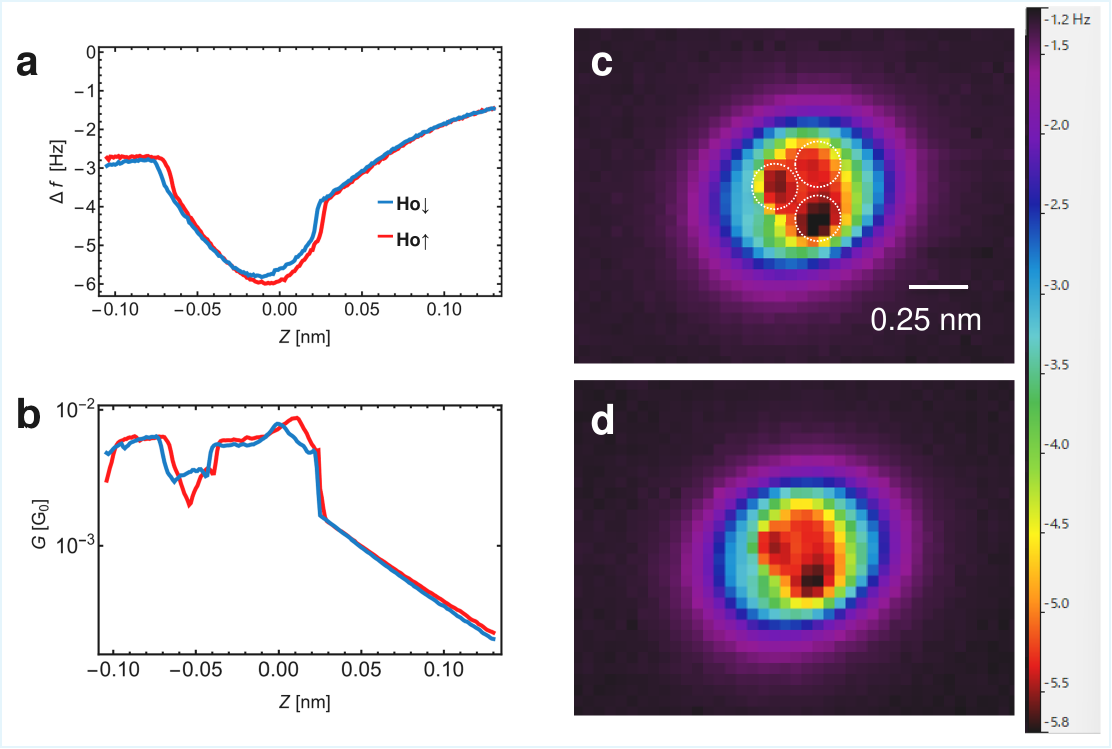}
\caption{\textbf{Imaging Ho adatom at small tip--sample distances.}
(a,b) The distance dependence of frequency shift and conductance obtained on top of the Ho$\uparrow$ ($\Delta$\textit{$f$}$_{\rm{Ho\uparrow}}$(\textit{$z$}), red solid curve) and Ho$\downarrow$ ($\Delta$\textit{$f$}$_{\rm{Ho\downarrow}}$(\textit{$z$}), blue solid curve). Measurement conditions: $V = \SI{+200}{\micro\volt}$ and $B = \SI{+3.0}{\tesla}$. Conductance is at the lowest turnaround point of the tip oscillation cycle obtained from the average tunneling current. (c,d) $\Delta$\textit{$f$} images of Ho$\uparrow$ and Ho$\downarrow$. Imaging parameter: constant height mode. $z = \SI{0}{\nm}$, $V = \SI{+200}{\micro\volt}$ and $B = \SI{+3.0}{\tesla}$. White dotted circles in (c) indicate the three local minima of $\Delta$\textit{$f$} in the Ho adatom.}\label{fig:FigS12}
\end{figure*}
\clearpage

\clearpage 
\clearpage
\begin{figure*}
\centering
\includegraphics[width=0.6\linewidth] {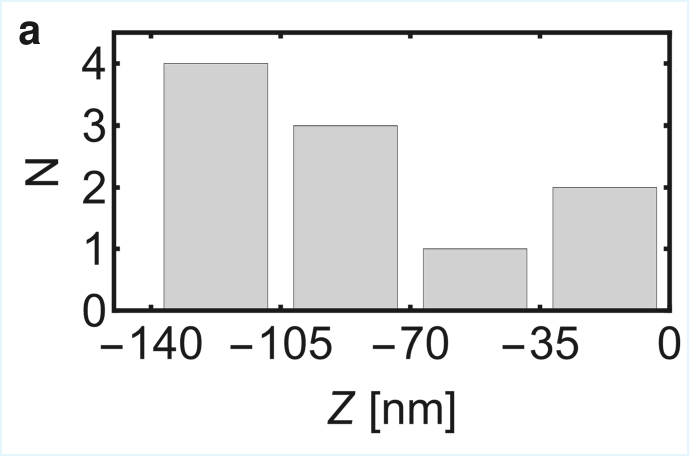}
\caption{\textbf{Force-based spin switching induced by different tip.}
(a) Histogram of tip–sample distance of spin switching using different tip shapes.}\label{fig:FigS13}
\end{figure*}
\clearpage

\clearpage
% \end{thebibliography}
\bibliography{ReferenceSI}

%%
%% End of file `elsarticle-template-num.tex'.

% Produces the bibliography via BibTeX.